\documentclass[aps,twocolumn,secnumarabic,balancelastpage,amsmath,amssymb,nofootinbib,hyperref=pdftex,prb]{revtex4-1}

\usepackage{graphics}
\usepackage[pdftex]{graphicx}
\graphicspath{{/Users/chrisreeg/Desktop/Doctoral_Research/My_Papers/RashbaTriplet/Figures/}}
\usepackage{longtable}
\usepackage{amsmath}
\usepackage{color,soul}

\makeatletter
\def\p@subsection{\thesection\,}
\makeatother
\newcommand{\bb}[1]{{\boldsymbol #1}}
\newcommand{\bo}[1]{{\bf #1}}

\newcommand{\G}{\mathcal{G}}
\newcommand{\F}{\mathcal{F}}
\newcommand{\uu}{{\uparrow\uparrow}}
\newcommand{\ud}{{\uparrow\downarrow}}
\newcommand{\du}{{\downarrow\uparrow}}
\newcommand{\dd}{{\downarrow\downarrow}}
\newcommand{\sgn}[1]{\text{sgn}(#1)}

\usepackage{hyperref}

\begin{document}

\title{Proximity-induced triplet superconductivity in Rashba materials}
\author{Christopher R. Reeg}
\author{Dmitrii L. Maslov}
\affiliation{Department of Physics, University of Florida, 
 P. O. Box 118440,
Gainesville, FL 32611-8440, USA}
\date{\today}
\begin{abstract}
We study a proximity junction between a conventional $s$-wave superconductor and a conductor with Rashba spin-orbit coupling, with a specific focus on the spin structure of the induced pairing amplitude. We find that spin-triplet pairing correlations are induced by spin-orbit coupling in both one- and two-dimensional systems due to the lifted spin degeneracy. Additionally, this induced triplet pairing has a component with an odd frequency dependence that is robust to disorder. Our predictions are based on the solutions of the exact Gor'kov equations and are beyond the quasiclassical approximation.
\end{abstract}

\maketitle

\section{Introduction} 
The generation of triplet superconducting correlations via proximity to a conventional superconductor (S) is a topic that has attracted a lot of attention recently in the context of superconducting spintronics, \cite{Eschrig:2011,Klose:2012,Quay:2013,Banerjee:2014,Banerjee:2014ly,Linder:2015} whereby triplet Cooper pairs with spin projection $\pm1$ play the same role as electron spins in conventional spintronics. \cite{Datta:1990,Zutic:2004} Most of the research in this area, both experimental and theoretical, has focused on proximity junctions involving ferromagnets (F), which induce triplet correlations by lifting the spin degeneracy (for a comprehensive review, see Refs.~\onlinecite{Buzdin:2005,Bergeret:2005}). The induced triplet correlations are also of fundamental interest because they have an odd frequency dependence; by the Pauli principle, these triplet states must be isotropic in momentum and therefore robust to disorder. \cite{Bergeret:2001,Volkov:2003,Yokoyama:2007}

Because spin-orbit coupling (SOC) also lifts the spin degeneracy, one might expect odd-frequency triplet pairing to emerge when a ferromagnet is replaced by a spin-orbit material as well. That the pairing symmetry of a bulk spin-orbit-coupled superconductor is known to be a mixture of singlet and triplet \cite{Edelstein:1989,Gorkov:2001,Frigeri:2004} only enhances these expectations. However, Liu et al. (Ref.~\onlinecite{Liu:2014}) showed, by solving the exact Bogoliubov-de Gennes (BdG) equations, that the proximity-induced pairing amplitude in a 1D metal with Rashba-type \cite{Bychkov:1984} SOC (R) has no triplet component. Previous studies of proximity junctions involving both SOC and ferromagnetism that were carried out to leading order in the quasiclassical approximation also did not find any triplet pairing induced by SOC. \cite{Bergeret:2013,*Bergeret:2014,*BergeretNote} Only very recently has the singlet-triplet mixing effect of SOC in such structures been noted by working beyond leading quasiclassical order. \cite{Bergeret:2015,Konschelle:2015} However, because the induced triplet pairing only shows up to second order, this is expected to be a very weak effect in any materials for which the quasiclassical methods are applicable. Spin-orbit scattering confined to the interface has been shown to generate triplet pairing in 3D ballistic superconductor/normal-metal junctions; however, this triplet component is anisotropic in momentum and therefore sensitive to disorder. \cite{Edelstein:2003} To date, odd-frequency triplet pairing induced by SOC in the proximity effect has not been explicitly investigated.

In this paper, we show by directly solving the fully quantum-mechanical Gor'kov equations that spin-triplet superconducting correlations are induced by Rashba SOC in both 1D and 2D proximity junctions. However, we find that the induced triplet component in 1D vanishes when integrated over the momentum; this result is in agreement with Ref.~\onlinecite{Liu:2014}. In 2D, we show that the induced triplet amplitude has an odd-frequency component that is isotropic in momentum. In agreement with Ref.~\onlinecite{Bergeret:2013,*Bergeret:2014}, we also find that the triplet pairing induced by SOC vanishes to leading order in the quasiclassical approximation. For this reason, our results are most relevant to materials that have a spin-orbit energy scale comparable to the Fermi energy, \cite{Rashba:2012} when no quasiclassical expansion can be made. Examples include  the surface states of noble metals \cite{LaShell:1996} and semi-metallic bismuth, \cite{Koroteev:2004} InSb quantum wires, \cite{vanWeperen:2015} as well as the bulk \cite{Ishizaka:2011} and surface \cite{Eremeev:2012} states of the bismuth tellurohalides. The proximity effect in materials with strong SOC is also relevant to recent experiments probing the existence of Majorana fermions in semiconductor quantum wires. \cite{Mourik:2012}

The remainder of the paper is organized as follows. In Sec.~\ref{Andreev}, we demonstrate that the presence of triplet pairing and the spatial dependence of the induced pairing amplitude can be deduced by considering Andreev reflection processes. We review the well-known S/F proximity effect from this point of view in Sec.~\ref{S/F} before discussing the qualitative similarities and differences of the S/R proximity effect in Sec.~\ref{S/R}. In Sec.~\ref{GorkovSec}, we solve the Gor'kov equations in the S/R geometry to show that triplet pairing is induced via the proximity effect. Our explicit model is described is Sec.~\ref{model} and details of the calculation are given in Sec.~\ref{details}. Solution details for the superconductor and Rashba regions are given in Sec.~\ref{GorkovSec}\,\ref{Ssolutions} and Sec.~\ref{GorkovSec}\,\ref{Rsolutions}, respectively, while the enforcement of boundary conditions is discussed in Sec.~\ref{GorkovSec}\,\ref{boundaryconditions}. Results in 1D and 2D are discussed in Secs.~\ref{1D} and \ref{2D}. In Sec.~\ref{quasiclassics}, we show that the induced triplet pairing amplitude vanishes to leading order in the quasiclassical limit. Our conclusions are given in Sec.~\ref{Conclusions}.

\begin{figure*}[t!]
\includegraphics[width=\linewidth]{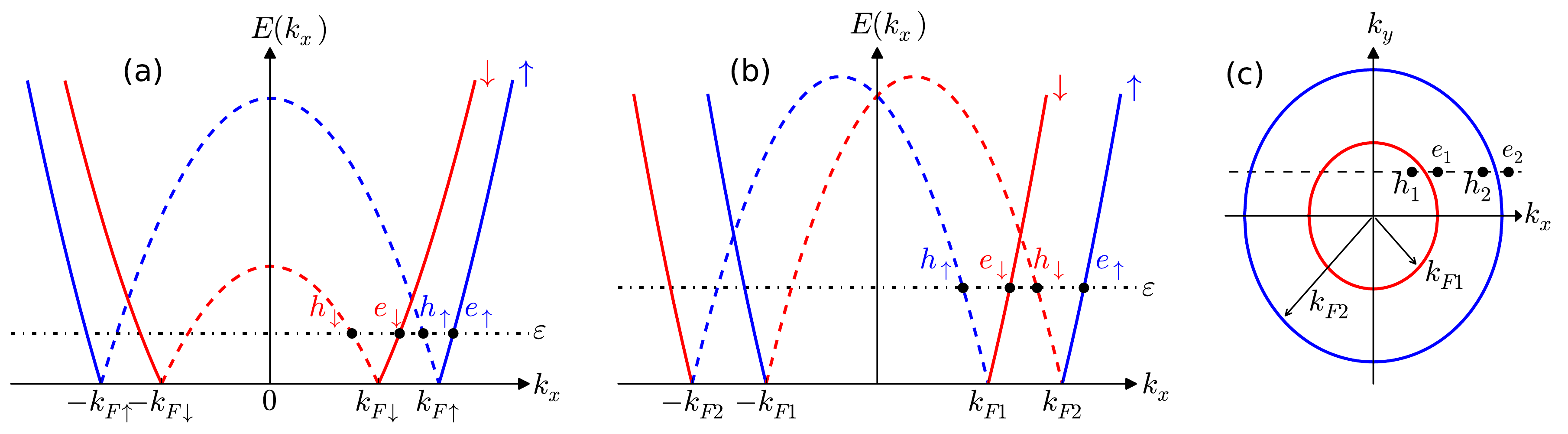}
\caption{\label{AndreevReflection} 
(Color online) Bogoliubov-de Gennes excitation spectra of (a) a 1D ferromagnet and (b) a 1D Rashba metal with Rashba vector $\bo{g}=(0,0,-ik_x)$. Colors denote different spin states, while solid (dashed) lines represent electron (hole) bands. An incident electron ($e_\uparrow$ or $e_\downarrow$) with energy $\varepsilon<\Delta$ is Andreev reflected as a hole of opposite spin ($h_\downarrow$ or $h_\uparrow$, respectively). The presence of a triplet pairing component with zero spin projection results from the lifted spin degeneracy. (c) Fermi surfaces of a 2D Rashba metal, with Rashba vector $\bo{g}=(ik_y,-ik_x,0)$, taking $E_{F}>0$. Because the two Rashba subbands are not associated with a definite spin, an incident electron ($e_1$ or $e_2$) with momentum $|k_y|<k_{F1}$ can be Andreev reflected into either subband ($h_1$ or $h_2$). The total pairing amplitude is a sum of oscillatory and nonoscillatory terms, resulting from interband and intraband Andreev reflection, respectively. Note: though they appear different, the 1D picture displayed in (b) and the $k_y=0$ limit of the 2D picture displayed in (c) are physically equivalent.}
\end{figure*}

\section{Triplet Pairing from Andreev Reflection} \label{Andreev}
The proximity effect can be understood qualitatively through Andreev reflection, whereby an electron (hole) with excitation energy $\varepsilon$ below the superconducting gap $\Delta$ is retroreflected as a hole (electron) with opposite spin, thus injecting a spin-singlet Cooper pair into the superconductor. \cite{Andreev:1964} While typically studied in the context of transport, \cite{Blonder:1982,deJong:1995,Yokoyama:2006,Sun:2015} Andreev reflection can also give insight into the type of pairing that is induced by the proximity effect. Due to the coherence between electrons and holes, it is these processes which allow the pairing amplitude to penetrate into the nonsuperconducting material.

By examining the eigenstates of the BdG equation, \cite{deGennes} we will show that SOC on its own is sufficient to induce triplet pairing in the proximity effect. For the sake of completeness, we will first review the case of triplet pairing in S/F junctions before extending the discussion to S/R junctions.

\subsection{Triplet pairing in S/F junctions} \label{S/F}
We first present a brief review of the proximity effect in ferromagnetic materials, following qualitative arguments similar to those given in Ref.~\onlinecite{Demler:1997}. In a ferromagnet, the BdG equation reads
\begin{equation} \label{BdGF}
\left(\begin{array}{cc} {\hat{\mathcal{H}}}_0-J\hat\sigma_z & 0 \\ 0 & -\hat{\mathcal{H}}_0+J\hat\sigma_z\end{array}\right)\psi(x,\varepsilon)=\varepsilon\psi(x,\varepsilon),
\end{equation}
where $\hat{\mathcal{H}}_0=(-\partial_x^2+k_\parallel^2)/2m-E_F$, with $k_\parallel$ being the conserved momentum along the S/F interface, $J$ is the ferromagnetic exchange field, $\hat\sigma_z$ is a Pauli matrix (we denote $2\times2$ matrices in spin space by a hat), and $\psi(x,\varepsilon)$ is a spinor wave function in Nambu $\otimes$ spin space. The 1D BdG excitation spectrum, containing spin-split bands with distinct Fermi momenta $k_{F\uparrow(\downarrow)}=\sqrt{2m(E_F\pm J)}\approx k_F\pm J/v_F$, is shown in Fig.~\ref{AndreevReflection}(a).

Because an incident electron [denoted by $e_\uparrow$ or $e_\downarrow$ in Fig.~\ref{AndreevReflection}(a)] must be Andreev reflected as a hole of opposite spin ($h_\downarrow$ or $h_\uparrow$, respectively), the resulting Cooper pairs in a 1D ferromagnet acquire a finite center-of-mass momentum $Q=k_{F\uparrow}-k_{F\downarrow}\approx 2h/v_F$, thus causing the Cooper pairing amplitude to oscillate in space with a period $\pi v_F/h$. [For arbitrary incidence, as is possible in higher dimensions, the center-of-mass momentum is modified to $Q=2J/v_F\cos\theta$, where $\theta$ is the angle the momentum makes with the interface normal.] The oscillations of the pairing amplitude are a direct consequence of the broken time-reversal symmetry, which implies that $E_{\bo{k}\uparrow}\neq E_{-\bo{k}\downarrow}$. As a result, paired states with opposite spins must have a finite total momentum. These oscillations are responsible for several interesting phenomena, including the possibility of a $\pi$ phase shift of the current-phase relation in S/F/S Josephson junctions, as well as nonmonotonic variations, as a function of the thickness of the ferromagnetic layer, of both the density of states and the superconducting critical temperature of S/F multilayers (see Refs.~\onlinecite{Buzdin:2005,Bergeret:2005} and the references therein for more detail).

As is more relevant to the current work, the lifting of the spin degeneracy (in this case by the exchange field) also results in the generation of a triplet pairing component that has zero spin projection. This is true for any material in which spin is an eigenstate of the Hamiltonian and can be seen particularly easily in 1D junctions by considering the contributions of the $e_\uparrow\to h_\downarrow$ and $e_\downarrow\to h_\uparrow$ Andreev reflection processes to the pairing amplitude. In terms of the BdG wave functions $u_\alpha(x,\varepsilon)$ (electrons) and $v_\alpha(x,\varepsilon)$ (holes), the triplet component of the pairing amplitude with zero spin projection is
\begin{equation}
\begin{aligned}
\F^\dagger_t(x,x',\varepsilon)&=\frac{1}{2}\left[\F_\ud^\dagger(x,x',\varepsilon)+\F_\du^\dagger(x,x',\varepsilon)\right] \\
	&\sim v_\uparrow(x,\varepsilon)u^*_\downarrow(x',\varepsilon)+v_\downarrow(x,\varepsilon)u^*_\uparrow(x',\varepsilon).
\end{aligned}
\end{equation}
The first term results from the $e_\downarrow\to h_\uparrow$ process, while the second term results from the $e_\uparrow\to h_\downarrow$ process. While the most general case is treated in Appendix~\ref{appendix}, we can express the triplet pairing amplitude in the case where the electrons are perfectly Andreev reflected by
\begin{equation}
\F_t^\dagger(x,x',\varepsilon)\sim a_\uparrow(\varepsilon)e^{ik_{\uparrow h}^Lx}e^{-ik_{\downarrow e}^Rx'}+a_\downarrow(\varepsilon)e^{ik_{\downarrow h}^Lx}e^{-ik_{\uparrow e}^Rx'},
\end{equation}
where $k^{R(L)}_{\uparrow(\downarrow)e(h)}$ is the momentum of a right-moving (left-moving) spin-up (spin-down) electron (hole) and $a_{\uparrow(\downarrow)}(\varepsilon)$ is the amplitude for Andreev reflection from a spin-down (spin-up) electron to a spin-up (spin-down) hole. As shown in Appendix~\ref{appendix}, the Andreev reflection amplitudes are related by $a_\uparrow(\varepsilon)=-a_\downarrow(\varepsilon)$. Therefore,
\begin{equation}
\F_t^\dagger(x,x',\varepsilon)\sim a_\uparrow(\varepsilon)\left(e^{ik_{\uparrow h}^Lx}e^{-ik_{\downarrow e}^Rx'}-e^{ik_{\downarrow h}^Lx}e^{-ik_{\uparrow e}^Rx'}\right).
\end{equation}
Because this quantity can only vanish when $k_\uparrow=k_\downarrow$, any lifting of the spin degeneracy will result in a nonzero triplet pairing amplitude.

\subsection{Triplet pairing in S/R junctions} \label{S/R}
While the proximity effect in S/R junctions is very similar to that in S/F junctions, there are several important qualitative differences we will now discuss. In the presence of SOC, the BdG equation is given by
\begin{equation} \label{BdG}
\left(\begin{array}{cc} \hat{\mathcal{H}}_0-i\lambda\bo{g}\cdot\hat{\bb{\sigma}} & 0 \\
	0 & -\hat{\mathcal{H}}_0-i\lambda\bo{g}\cdot\hat{\bb{\sigma}}^* \end{array}\right)\psi(x,\varepsilon)=\varepsilon\psi(x,\varepsilon),
\end{equation}
where $\lambda$ is the SOC constant and $\bo{g}$ is the spin-orbit vector, which satisfies $\bo{g}(\bo{k})=-\bo{g}(-\bo{k})$.

In a strictly 1D system, where there is no well-defined spin quantization axis, the two representations of the Rashba vector $\bo{g}=(0,-ik_x,0)$ and $\bo{g}=(0,0,-ik_x)$ are physically equivalent. To discuss Andreev reflection in the 1D case, we choose the latter representation, so that spin remains an eigenstate of the BdG Hamiltonian. The BdG excitation spectrum, which contains two Rashba subbands with Fermi momenta $k_{F1(2)}=\sqrt{(m\lambda)^2+2mE_F}\mp m\lambda$, is shown in Fig.~\ref{AndreevReflection}(b). While we take $E_F>0$ in Fig.~\ref{AndreevReflection}(b), so that the Fermi energy lies above the Dirac point, the following physical arguments are equally valid if $E_F<0$. Because each Rashba subband can be associated with a definite spin, an electron incident on the S/R interface must be Andreev reflected as a hole with nearly equal momentum [close to $k_{F1}$ for the $e_\downarrow\to h_\uparrow$ process and to $k_{F2}$ for the $e_\uparrow\to h_\downarrow$ process, as labeled in Fig.~\ref{AndreevReflection}(b)]. Due to the momentum matching between the incident electron and reflected hole, only zero-momentum Cooper pairs are formed within R and the induced pairing amplitude does not oscillate as a function of the center-of-mass coordinate of the Cooper pair. The spatial dependence of the pairing amplitude can also be inferred from the preservation of time-reversal symmetry; because $E_{k\uparrow}=E_{-k\downarrow}$, paired states of opposite spin have zero total momentum. As shown in Sec.~\ref{S/F}, the presence of a triplet pairing component again follows from the lifted spin degeneracy.

Unlike in the ferromagnetic case, the proximity effect in 2D S/R junctions is qualitatively different than in 1D. In 2D, the Rashba spin-orbit vector lies in the plane: $\bo{g}=(ik_y,-ik_x,0)$. Spin-orbit coupling again splits the Fermi surface into two Rashba subbands, see Fig.~\ref{AndreevReflection}(c), but the crucial difference compared to 1D is that these subbands can no longer be associated with a definite spin. Therefore, for the case where $E_{F}>0$ and $|k_y|<k_{F1}$, as illustrated in Fig.~\ref{AndreevReflection}(c), interband reflections ($e_1\to h_2$ or $e_2\to h_1$) are allowed. Because the momentum of an Andreev reflected hole within the opposite band differs from that of the incident electron, the Cooper pairs that are formed by processes of this type have a finite center-of-mass momentum. The induced pairing amplitude in two dimensions is thus a sum of oscillatory (from interband processes) and nonoscillatory (from intraband processes) terms. The proximity effect is qualitatively similar when $E_F<0$; even though only a single Rashba subband is occupied, the inner and outer radii of the annular Fermi surface play the same role as the two distinct Fermi momenta in Fig.~\ref{AndreevReflection}(c).

\vspace*{0.1in} 

\section{Solutions of Gor'kov equations} \label{GorkovSec}
\subsection{Model and Equations} \label{model}
\begin{figure}[b!]
\includegraphics[width=\linewidth]{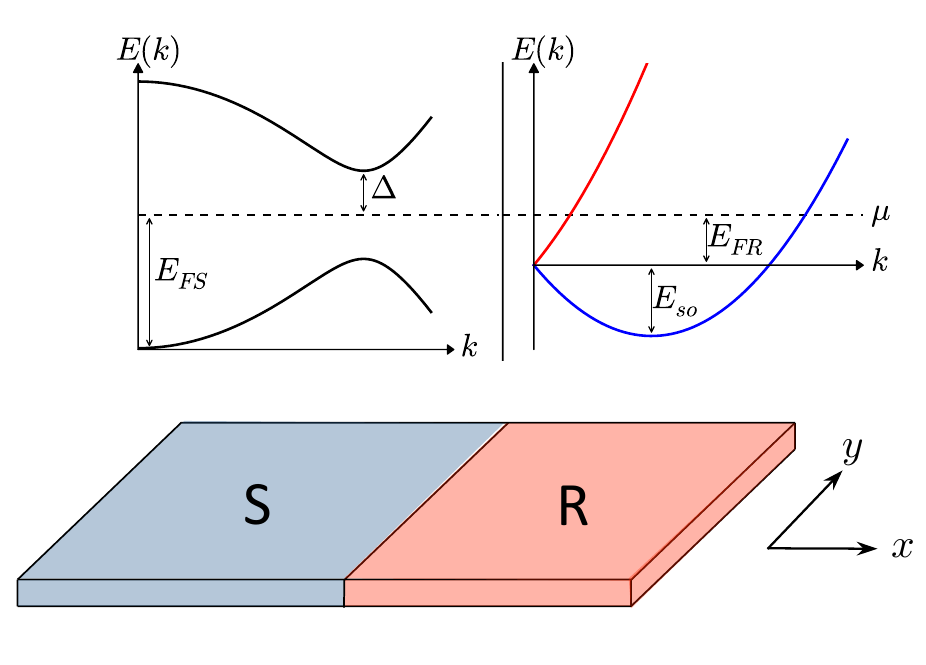}
\caption{\label{setup} 
Geometry of the S/R proximity effect, together with the band structure of each material. System is taken to be infinite in the $y$ direction in 2D, while 1D corresponds to the single-channel limit $k_y=0$.
}
\end{figure}

We will now show how the physics described in Sec.~\ref{S/R} follows from the microscopic theory. We consider a two-dimensional model of a S/R proximity junction (Fig.~\ref{setup}) and allow the mass $m(x)$, the SOC constant $\lambda(x)$, the Fermi energy $E_F(x)$, and the pairing potential $\Delta(x)$ to vary in a stepwise manner across the S/R interface. Specifically, we take $m(x)=m_R\theta(x)+m_S\theta(-x)$, $E_{F}(x)=E_{FR}\theta(x)+E_{FS}\theta(-x)$, $\lambda(x)=\lambda\theta(x)$, and $\hat\Delta(x)=\Delta\theta(-x)i\hat{\sigma}_y$. The model can be represented by an explicitly Hermitian Hamiltonian,
\begin{equation} \label{H}
\begin{aligned}
H&=\int d^2x\biggl\{\psi^\dagger(x)\biggl[-\frac{1}{2}\partial_x\left(\frac{1}{m(x)}\partial_x\right)+\frac{k_y^2}{2m(x)}-E_F(x) \\
	&-\frac{i}{2}\hat{\sigma}_y\bigl\{\lambda(x)\partial_x+\partial_x[\lambda(x)]\bigr\}-\hat\sigma_x\lambda(x)k_y\biggr]\psi(x) \\
	&+\frac{1}{2}\left(\psi^\dagger(x)\hat\Delta(x)[\psi^\dagger(x)]^T+[\psi(x)]^T\hat\Delta^\dagger(x)\psi(x)\right)\biggr\}.
\end{aligned}
\end{equation}
Interfacial scattering is incorporated through a mismatch in Fermi velocities/momenta across the S/R interface. The Gor'kov equations in this model are given by \cite{MineevBook} 
\vfill
\begin{widetext}
\begin{subequations} \label{Gorkov}
\begin{gather}
\left[i\omega+\frac{1}{2}\partial_x\left(\frac{1}{m(x)}\partial_x\right)-\frac{k_y^2}{2m(x)}+E_F(x)-\frac{i}{2}\hat\sigma_y\bigl\{\lambda(x)\partial_x+\partial_x[\lambda(x)]\bigr\}-\hat\sigma_x\lambda(x)k_y\right]\hat{\G}_{\omega,k_y}(x,x')+\hat\Delta(x)\hat\F^\dagger_{\omega,k_y}(x,x')=\delta(x-x'), \\
\left[-i\omega+\frac{1}{2}\partial_x\left(\frac{1}{m(x)}\partial_x\right)-\frac{k_y^2}{2m(x)}+E_F(x)-\frac{i}{2}\hat\sigma_y\bigl\{\lambda(x)\partial_x+\partial_x[\lambda(x)]\bigr\}+\hat\sigma_x\lambda(x)k_y\right]\hat\F^\dagger_{\omega,k_y}(x,x')-\hat\Delta^\dagger(x)\hat\G_{\omega,k_y}(x,x')=0.
\end{gather}
\end{subequations}
\end{widetext}
We solve the fully quantum-mechanical Gor'kov equations rather than the quasiclassical Eilenberger equations so that we can treat the limit of strong SOC, where the splitting of Rashba subbands must be taken into account.

Even without performing a detailed calculation, it is evident from Eq.~(\ref{Gorkov}b) that SOC generates triplet pairing in the proximity region. If we parameterize the pairing amplitude by
\begin{equation} \label{decomp}
\hat\F^{\dagger}=(s+\bo{d}\cdot\hat{\bb{\sigma}})i\hat\sigma_y,
\end{equation}
then the four coupled Gor'kov equations describing the pairing induced in R can be written out explicitly as
\begin{equation} \label{induced}
\begin{aligned}
&\left(a_R^{*2}+\partial_x^2\right) s+2m_R\lambda(k_yd_x-i\partial_xd_y)=0, \\
&\left(a_R^{*2}+\partial_x^2\right) d_x+2m_R\lambda(k_ys+\partial_xd_z)=0, \\
&\left(a_R^{*2}+\partial_x^2\right) d_y-2m_Ri\lambda(k_yd_z-\partial_xs)=0, \\
&\left(a_R^{*2}+\partial_x^2\right) d_z+2m_R\lambda(ik_yd_y-\partial_xd_x)=0,
\end{aligned}
\end{equation}
where we define $a_{R(S)}^2=2m_{R(S)}[i\omega+E_{FR(S)}]-k_y^2$. In 2D ($k_y\neq0$), all three triplet components are coupled to the singlet component through SOC, whereas only the component $d_y$ is coupled to $s$ in 1D ($k_y=0$).

\subsection{Calculation Details} \label{details}
We now proceed with detailed solutions of Eq.~(\ref{Gorkov}) in both the superconductor and Rashba metal. Because we are ultimately interested in the pairing amplitude induced in R, we fix $x'>0$. In the context of Eq.~(\ref{Gorkov}), this choice for $x'$ simply ensures that the delta-function source term on the right-hand side appears only when solving for the Green's functions in R.

\subsubsection{Solutions in superconductor} \label{Ssolutions}
We begin by solving the Gor'kov equations within S, which can be expressed as
\begin{subequations}
\begin{align}
\bigl(a_S^2+\partial_x^2\bigr)\hat\G^S_{\omega,k_y}(x,x')+2m_S\Delta i\hat\sigma_y\hat\F^{\dagger S}_{\omega,k_y}(x,x')&=0, \label{GorkovS1} \\
\bigl(a_S^{*2}+\partial_x^2\bigr)\hat\F^{\dagger S}_{\omega,k_y}(x,x')+2m_S\Delta i\hat\sigma_y\hat\G^{S}_{\omega,k_y}(x,x')&=0. \label{GorkovS2}
\end{align}
\end{subequations}
We denote Green's functions in the superconductor by
\begin{subequations}
\begin{align}
\hat\G^S_{\omega,k_y}(x,x')&=\hat\G_{\omega,k_y}(x<0,x'), \\
\hat\F^{\dagger S}_{\omega,k_y}(x,x')&=\hat\F^\dagger_{\omega,k_y}(x<0,x').
\end{align}
\end{subequations}
Solving Eq.~(\ref{GorkovS2}) for $\hat{\G}$, we obtain
\begin{equation} \label{GFrelation}
\hat\G_{\omega,k_y}^S(x,x')=i\hat\sigma_y\left(\frac{a_S^{*2}+\partial_x^2}{2m_S\Delta}\right)\hat\F^{\dagger S}_{\omega,k_y}(x,x').
\end{equation}
Substituting this expression for $\hat\G$ into Eq.~(\ref{GorkovS1}), we obtain a fourth-order equation describing $\hat{\F}^{\dagger}$,
\begin{equation} \label{GorkovSF}
\bigl[(a_S^2+\partial_x^2)(a_S^{*2}+\partial_x^2)+4m_S^2\Delta^2\bigr]\hat\F^{\dagger S}_{\omega,k_y}(x,x')=0.
\end{equation}
Solving Eq.~(\ref{GorkovSF}) and keeping only those solutions that decay in the limit $x\to-\infty$, we obtain
\begin{equation} \label{SsolutionF}
\hat\F^{\dagger S}_{\omega,k_y}(x,x')=\hat{c}_1(x')e^{-ipx}+\hat{c}_5(x')e^{ip^*x}, \\
\end{equation}
where $p^2=2m_S(i\Omega+E_{FS})-k_y^2$ and $\Omega^2=\omega^2+\Delta^2$. Here and throughout the rest of the calculation, we choose the branch cut of the square root function to lie along the negative real axis. Note that each coefficient $\hat{c}_i$ is a $2\times2$ matrix in spin space and thus contains four unknown quantities,
\begin{equation}
\hat{c}_1(x')=\left(\begin{array}{cc} c_1(x') & c_2(x') \\ c_3(x') & c_4(x') \end{array}\right).
\end{equation}
Substituting solution (\ref{SsolutionF}) into Eq.~(\ref{GFrelation}) gives the normal Green's function,
\begin{equation} \label{SsolutionG}
\begin{aligned}
\hat{\G}^S_{\omega,k_y}(x,x')&=-i\hat\sigma_y\left[\hat{c}_1(x')e^{-ipx}e^{i\eta}+\hat{c}_5(x')e^{ip^*x}e^{-i\eta}\right],
\end{aligned}
\end{equation}
where we define $\eta=\cos^{-1}(i\omega/\Delta)$.

\subsubsection{Solutions in Rashba metal} \label{Rsolutions}
The Gor'kov equation describing the normal Green's function in R is given by
\begin{equation} \label{GorkovR}
\bigl[a_R^2+\partial_x^2-2m_R\lambda(i\hat\sigma_y\partial_x+\hat\sigma_xk_y)\bigr]\hat\G^R_{\omega,k_y}(x,x')=\delta(x-x').
\end{equation}
(Note that $\hat{\G}^R$ denotes the Matsubara Green's function in the Rashba metal, 
\begin{equation}
\hat{\G}^R_{\omega,k_y}(x,x')=\hat\G_{\omega,k_y}(x>0,x'),
\end{equation}
and should not be confused for a retarded Green's function.) Equation~(\ref{GorkovR}) consists of four equations describing the four spin components of the matrix $\hat\G$. This system can be easily solved by considering two equations at a time; for example (we suppress explicit reference to the dependence of the Green's function on $\omega$ and $k_y$ here),
\begin{subequations} \label{subeq1}
\begin{gather}
(a_R^2+\partial_x^2)\G^R_\uu(x,x')-2m_R\lambda(k_y+\partial_x)\G^R_\du(x,x')=\delta(x-x'), \\
(a_R^2+\partial_x^2)\G^R_\du(x,x')-2m_R\lambda(k_y-\partial_x)\G^R_\uu(x,x')=0.
\end{gather}
\end{subequations}
The Green's functions can be uniquely expressed as the sum of a particular solution to Eq.~(\ref{subeq1}) and the general solution to the corresponding homogeneous system of equations. 

The particular solution to Eq.~(\ref{subeq1}) is equal to the Green's function of a bulk Rashba metal, which we denote by $\G^{(0)R}_{\alpha\beta}(x-x')$, and is a function of only the difference $x-x'$. This function can be obtained by Fourier transforming to momentum space,
\begin{subequations} \label{subeq2}
\begin{gather}
(a_R^2-k_x^2)\G^{(0)R}_\uu(k_x)-2m_R\lambda(k_y+ik_x)\G^{(0)R}_\du(k_x)=1, \\
(a_R^2-k_x^2)\G^{(0)R}_\du(k_x)-2m_R\lambda(k_y-ik_x)\G^{(0)R}_\uu(k_x)=0.
\end{gather}
\end{subequations}
Solving the algebraic system (\ref{subeq2}), we find
\begin{subequations} \label{momsol}
\begin{align}
\G_\uu^{(0)R}(k_x)&=\frac{2m_R(a_R^2-k_x^2)}{(a_R^2-k_x^2)^2-4m_R^2\lambda^2(k_x^2+k_y^2)}, \\
\G_\du^{(0)R}(k_x)&=\frac{4m_R^2\lambda(k_y-ik_x)}{(a_R^2-k_x^2)^2-4m_R^2\lambda^2(k_x^2+k_y^2)}.
\end{align}
\end{subequations}
To obtain the Green's function in coordinate space, we perform an inverse Fourier transform,
\begin{equation}
\G^{(0)R}_{\alpha\beta}(x-x')=\int\frac{dk_x}{2\pi}\G_{\alpha\beta}^{(0)R}(k_x)e^{ik_x(x-x')}.
\end{equation}
The integrals evaluate to
\begin{subequations} \label{bulksol}
\begin{align}
\G^{(0)R}_\uu(x-x')&=-\frac{im_R}{k_1^{1D}+k_2^{1D}}\biggl[\frac{s_1k_1^{1D}}{k_1}e^{ik_1s_1|x-x'|} \\
	\nonumber &+\frac{s_2k_2^{1D}}{k_2}e^{ik_2s_2|x-x'|}\biggr], \\
\nonumber \G^{(0)R}_\du(x-x')&=\frac{im_R}{k_1^{1D}+k_2^{1D}}\biggl\{\left[i\,\sgn{x-x'}-\frac{s_1k_y}{k_1}\right] \\
	\nonumber &\times e^{ik_1s_1|x-x'|}-\left[i\,\sgn{x-x'}-\frac{s_2k_y}{k_2}\right] \\
	&\times e^{ik_2s_2|x-x'|}\biggr\}.
\end{align}
\end{subequations}
Whereas the states in the superconductor are characterized by a single momentum $p$, in the Rashba metal we must define two different momenta due to the band splitting by SOC,
\begin{equation}
k_{1(2)}^2=\bigl[k_{1(2)}^{1D}\bigr]^2-k_y^2,
\end{equation}
where
\begin{equation}
k_{1(2)}^{1D}=\sqrt{m_R^2\lambda^2+2m_R(i\omega+E_{FR})}\mp m_R\lambda
\end{equation}
are the corresponding momenta in 1D. We also denote $s_\alpha=\text{sgn}[\text{Im}(k_\alpha)]$, which evaluates to
\begin{subequations}
\begin{align}
\label {s1} s_1&=\left\{\begin{array}{rcl}
	\sgn{\omega} & \text{if} & E_{FR}>0, \\
	-\sgn{\omega} & \text{if} & E_{FR}<0,\omega^2<-4E_{so}E_{FR}, \\
	\sgn{\omega} & \text{if} & E_{FR}<0,\omega^2>-4E_{so}E_{FR}, \end{array}\right.  \\
\label{s2} s_2&=\sgn{\omega}.
\end{align}
\end{subequations}
In Eq.~(\ref{s1}), we define the spin-orbit energy $E_{so}=m_R\lambda^2/2$ (see also Fig.~\ref{setup}). Based on the symmetry of Eq.~(\ref{GorkovR}), the remaining spin components of the bulk Rashba solution can be obtained directly from Eq.~(\ref{bulksol}) by noting that $\G_\uu\to\G_\dd$ and $\G_\du\to\G_\ud$ upon flipping the signs of both $\lambda$ and $k_y$. The full bulk solution is given by
\begin{equation} \label{Gsolbulk}
\begin{aligned}
&\hat{\G}^{(0)R}_{\omega,k_y}(x-x')=-\frac{im_Rs_1}{k_1^{1D}+k_2^{1D}}\biggl[\frac{k_1^{1D}}{k_1}-s_1\,\sgn{x-x'}\hat\sigma_y \\
	&\hspace*{0.2in}+\frac{k_y}{k_1}\hat\sigma_x\biggr]e^{ik_1s_1|x-x'|}-\frac{im_Rs_2}{k_1^{1D}+k_2^{1D}}\biggl[\frac{k_2^{1D}}{k_2} \\
	&\hspace*{0.2in}+s_2\,\sgn{x-x'}\hat\sigma_y-\frac{k_y}{k_2}\hat\sigma_x\biggr]e^{ik_2s_2|x-x'|}. \\
\end{aligned}
\end{equation}

We now seek solutions to the homogeneous system of equations
\begin{subequations} \label{subeq3}
\begin{gather}
(a_R^2+\partial_x^2)\G^R_\uu(x,x')-2m_R\lambda(k_y+\partial_x)\G^R_\du(x,x')=0, \\
(a_R^2+\partial_x^2)\G^R_\du(x,x')-2m_R\lambda(k_y-\partial_x)\G^R_\uu(x,x')=0.
\end{gather}
\end{subequations}
Because this is a linear system, it can easily be solved by matrix methods. We transform the system (\ref{subeq3}) into a first-order matrix equation,
\begin{equation} \label{mat}
\partial_x\bo{X}=\left(\begin{array}{cccc} 
	0 & 0 & 1 & 0 \\
	0 & 0 & 0 & 1 \\
	-a_R^2 & -2m_R\lambda k_y & 0 & -2m_R\lambda \\
	-2m_R\lambda k_y & -a_R^2 & 2m_R\lambda & 0 \end{array}\right)\bo{X},
\end{equation}
describing the vector $\bo{X}=[\G^R_\uu,\G^R_\du,\partial_x\G^R_\uu,\partial_x\G^R_\du]^T$. The four eigenvalues of the matrix in Eq.~(\ref{mat}) are $\pm ik_1$ and $\pm ik_2$, but we choose only the solutions that decay in the limit $x\to\infty$. The prefactors of the exponentials in the solution are determined by the eigenvectors. Thus, a full solution to the Gor'kov equations in the Rashba metal is given by
\begin{equation} \label{Gsol}
\begin{aligned}
\hat\G^R_{\omega,k_y}(x,x')&=\left(\begin{array}{cc}
	\frac{ik_1s_1+k_y}{k_1^{1D}}c_9(x') & c_{10}(x') \\
	c_9(x') & -\frac{ik_1s_1-k_y}{k_1^{1D}}c_{10}(x')
\end{array}\right)e^{ik_1s_1 x} \\
	&+\left(\begin{array}{cc}
	-\frac{ik_2s_2+k_y}{k_2^{1D}}c_{11}(x') & c_{12}(x') \\
	c_{11}(x') & \frac{ik_2s_2-k_y}{k_2^{1D}}c_{12}(x')
\end{array}\right)e^{ik_2s_2 x} \\
&+\hat\G^{(0)R}_{\omega,k_y}(x-x').
\end{aligned}
\end{equation}

Turning now to the solution for the pairing amplitude in the Rashba metal, the Gor'kov equations that we must solve are given by
\begin{equation} \label{GorkovR2}
\bigl[a_R^{*2}+\partial_x^2-2m_R\lambda(i\hat\sigma_y\partial_x-\hat\sigma_xk_y)\bigr]\hat\F^{\dagger R}_{\omega,k_y}(x,x')=0.
\end{equation}
Noting that Eq.~(\ref{GorkovR2}) becomes Eq.~(\ref{GorkovR}) upon flipping the signs of both $\omega$ and $k_y$ (save for the delta-function term), we obtain the pairing amplitude directly from Eq.~(\ref{Gsol}) by making these changes,
\begin{equation} \label{Fsol} 
\begin{aligned}
\hat\F^{\dagger R}_{\omega,k_y}(x,x')&=\left(\begin{array}{cc}
	-\frac{ik_1^*s_1+k_y}{k_1^{*1D}}c_{13}(x') & c_{14}(x') \\
	c_{13}(x') & \frac{ik_1^*s_1-k_y}{k_1^{*1D}}c_{14}(x')
\end{array}\right)e^{-ik_1^*s_1 x} \\
	&+\left(\begin{array}{cc}
	\frac{ik_2^*s_2+k_y}{k_2^{*1D}}c_{15}(x') & c_{16}(x') \\
	c_{15}(x') & -\frac{ik_2^*s_2-k_y}{k_2^{*1D}}c_{16}(x')
\end{array}\right)e^{-ik_2^*s_2 x}.
\end{aligned}
\end{equation}

\subsubsection{Enforcing boundary conditions} \label{boundaryconditions}
Now that we have obtained general solutions to the Gor'kov equations in both the superconductor and Rashba metal, the sixteen unknown coefficients must be determined by boundary conditions. The boundary conditions can be obtained by direct integration of Eq.~(\ref{Gorkov}) over a narrow region near $x=0$; they are
\begin{subequations} \label{BCs}
\begin{gather}
\hat{\G}^R_{\omega,k_y}(0,x')=\hat{\G}^S_{\omega,k_y}(0,x'), \\
\left(\frac{\partial_x}{m_R}-i\hat\sigma_y\lambda\right)\hat{\G}^R_{\omega,k_y}(0,x')=\frac{\partial_x}{m_S}\hat{\G}^S_{\omega,k_y}(0,x'),
\end{gather}
\end{subequations}
with the same boundary conditions applying for $\hat{\F}^\dagger$ as well. 

Because the bulk solution (\ref{Gsolbulk}) contains two linearly independent terms, we can separate the dependence of the coefficients on $x'$ by writing
\begin{equation} \label{coeffs}
c_i(x')=c_{i,1}e^{ik_1s_1 x'}+c_{i,2}e^{ik_2s_2 x'}.
\end{equation}
Each of the sixteen boundary conditions then separates into two linearly independent parts, thus giving a total of 32 boundary conditions that must be solved. The pairing amplitude in Eq.~(\ref{Fsol}) becomes
\begin{widetext} 
\begin{equation} \label{Fsol2} 
\begin{aligned} \hat\F^{\dagger R}_{\omega,k_y}(x,x')&=\left(\begin{array}{cc}
	-\frac{ik_1^*s_1+k_y}{k_1^{*1D}}c_{13,1} & c_{14,1} \\
	c_{13,1} & \frac{ik_1^*s_1-k_y}{k_1^{*1D}}c_{14,1}
\end{array}\right)e^{-ik_1^*s_1 x}e^{ik_1s_1x'}+\left(\begin{array}{cc}
	-\frac{ik_1^*s_1+k_y}{k_1^{*1D}}c_{13,2} & c_{14,2} \\
	c_{13,2} & \frac{ik_1^*s_1-k_y}{k_1^{*1D}}c_{14,2}
\end{array}\right)e^{-ik_1^*s_1 x}e^{ik_2s_2x'} \\
	&+\left(\begin{array}{cc}
	\frac{ik_2^*s_2+k_y}{k_2^{*1D}}c_{15,1} & c_{16,1} \\
	c_{15,1} & -\frac{ik_2^*s_2-k_y}{k_2^{*1D}}c_{16,1}
\end{array}\right)e^{-ik_2^*s_2 x}e^{ik_1s_1x'}+\left(\begin{array}{cc}
	\frac{ik_2^*s_2+k_y}{k_2^{*1D}}c_{15,2} & c_{16,2} \\
	c_{15,2} & -\frac{ik_2^*s_2-k_y}{k_2^{*1D}}c_{16,2}
\end{array}\right)e^{-ik_2^*s_2 x}e^{ik_2s_2x'}.
\end{aligned}
\end{equation}
\end{widetext}
Due to the matched momenta in the exponentials of the first and last terms of Eq.~(\ref{Fsol2}), these terms correspond to Cooper pairs with zero net momentum; these terms do not oscillate as a function of the center-of-mass coordinate of the pair $(x+x')/2$. Conversely, the terms with mismatched momenta correspond to Cooper pairs with a finite momentum; these terms are oscillatory.

To compactify our notation, we can express the singlet and triplet parts of the induced pairing amplitude, as defined in Eq.~(\ref{decomp}), as
\begin{equation} \label{amplitude}
(s,\bo{d})=\sum_{\alpha,\beta=1}^2[f^0_{\alpha\beta}(\omega,k_y),\bo{f}_{\alpha\beta}(\omega,k_y)]e^{-ik_\alpha^*s_\alpha x}e^{ik_\beta s_\beta x'}.
\end{equation}
The sums in Eq.~(\ref{amplitude}) run over the two Rashba subbands and $(f^0,\bo{f})$ describes a four-vector of $2\times2$ matrices that can be directly related to the spatial dependence of the induced pairing amplitude. The four elements of the newly defined $f^i$ matrices correspond to the four terms of Eq.~(\ref{Fsol2}); for example, the upper diagonal element of $f^0$ is given by the singlet component of the matrix in the first term of Eq.~(\ref{Fsol2}), $f^0_{11}=(c_{14,1}-c_{13,1})/2$. For this reason, we associate the diagonal elements of the $f^i$ matrices with nonoscillatory terms of the pairing amplitude and off-diagonal elements with oscillatory terms.

\begin{figure*}[t!]
\includegraphics[width=\linewidth]{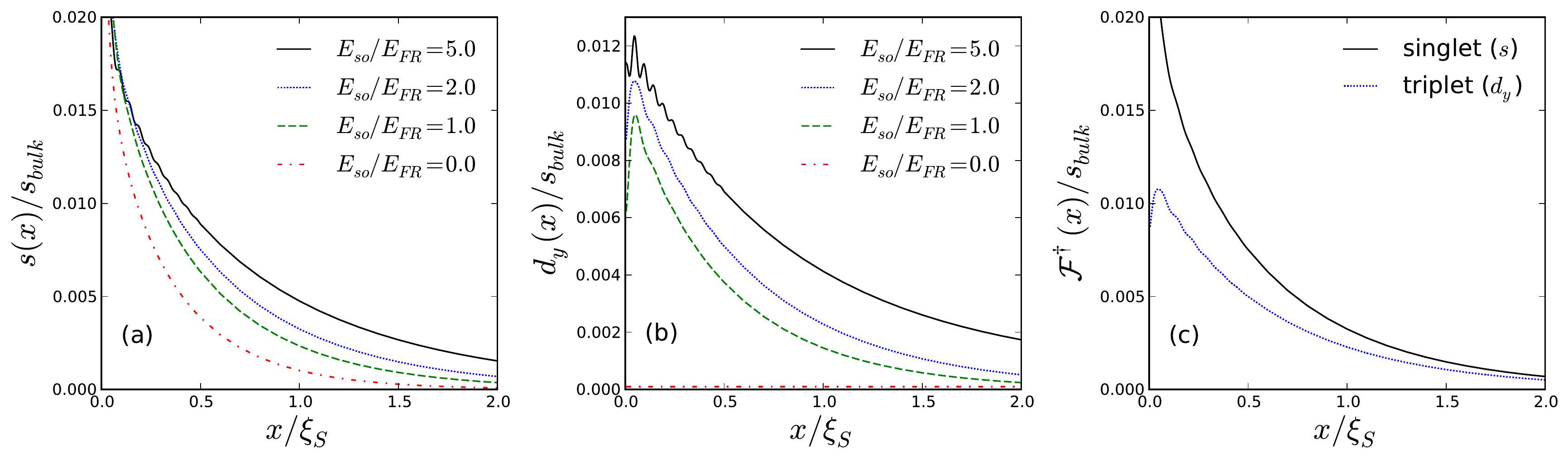}
\caption{\label{comparison} (Color online) Spatial dependence of proximity-induced (a) singlet and (b) triplet pairing amplitudes, as defined in Eq.~(\ref{def}), in a 2D Rashba proximity junction for various strengths of SOC ($\xi_S=v_F/\Delta$ is the coherence length of the superconductor). Pairing amplitude is plotted in units of its value in the bulk of the superconductor. (c) Quantitative comparison between induced singlet and triplet amplitudes for $E_{so}/E_{FR}=2$ [dotted curves from (a) and (b)], which is a reasonable choice for BiTeI. \cite{VanGennep:2014} All pairing amplitudes are plotted with $\omega/\Delta=0.8$, $m_R/m_S=0.1$, $E_{FR}/E_{FS}=0.05$, and $E_{FS}/\Delta=200$. By the Pauli principle, the triplet component has an odd frequency dependence.
}
\end{figure*}

\subsection{Pairing Amplitude in 1D} \label{1D}
Solving the boundary conditions in 1D, we obtain ${f}^x= f^z=0$, while $ f^y\propto\sigma_z$ and $ f^0\propto\sigma_0$. Consistent with our previous discussion of Andreev reflection, we find that the pairing amplitude is a mixture of singlet and triplet components and does not oscillate with $(x+x')/2$. In order to simplify the analytic result, we assume that the Fermi energy and the spin-orbit energy are the largest energy scales, so that $\omega,\Delta\ll E_{so},E_{FR(S)}$. This allows us to expand
\begin{subequations}
\begin{align}
p&=p_F+i\Omega/v_F, \\
k_1&=(k_{F1}+i\omega/v_R)\sgn{E_{FR}}, \\
k_2&=k_{F2}+i\omega/v_R,
\end{align}
\end{subequations}
where $v_F=\sqrt{2E_{FS}/m_S}=p_F/m_S$ and $v_R=\sqrt{\lambda^2+2E_{FR}/m_R}$ are the Fermi velocities of the superconductor and Rashba metal, respectively. We also note that, in this limit, $s_1=\sgn\omega\sgn{E_{FR}}$. The sole nonzero triplet component can then be expressed as
\begin{equation} \label{triplet}
\begin{aligned}
d_y(x,x')&=-\frac{v_F\Delta\,\sgn\omega}{2v_Fv_R|\omega|+(v_F^2+v_R^2)\Omega\,\sgn{E_{FR}}}e^{-\frac{|\omega|}{v_R}(x+x')} \\
	&\times \left(e^{-ik_{F1}\,\sgn\omega(x-x')}-e^{-ik_{F2}\,\sgn\omega(x-x')}\right).
\end{aligned}
\end{equation}
As is customary, we can also express the nonlocal pairing amplitude in terms of the center-of-mass coordinate and the momentum of the relative motion. In this mixed representation ($x$ now denotes the center-of-mass coordinate), the triplet pairing amplitude when $E_{FR}>0$ is given by
\begin{equation} \label{triplet2}
\begin{aligned}
d_y(x,k)&=\frac{2\pi v_Fv_R\Delta\,\sgn k}{2v_Fv_R|\omega|+(v_F^2+v_R^2)\Omega}e^{-2|\omega|x/v_R} \\
	&\times\bigl[\delta(\xi_1)-\delta(\xi_2)\bigr]\theta(-k\omega),
\end{aligned}
\end{equation}
where $\xi_{1(2)}=v_R(|k|-k_{F1(2)})$; a similar expression can be written if $E_{FR}<0$. In both cases, the triplet component in Eq.~(\ref{triplet2}) consists of terms localized to each of the two split Fermi surfaces. However, this triplet component vanishes if integrated over the momentum, as the integrated triplet amplitude is proportional to the difference in the densities of states on the two Rashba subbands,
\begin{equation}
d_y(x)=\int \frac{dk}{2\pi}d_y(x,k)\propto N_1(E_{FR})-N_2(E_{FR}).
\end{equation} 
Since the subband densities of states in 1D are equal regardless of the position of the Fermi level, $N_1(E_{FR})=N_2(E_{FR})=1/\pi v_R$, $d_y(x)$ vanishes. This result is consistent with that of Ref.~\onlinecite{Liu:2014}, which found no induced local (equivalently, momentum-integrated) triplet pairing amplitude in 1D. [Additional Rashba coupling arising from a lateral confining potential in quasi-1D wires makes the velocities of spin-split subbands, and thus the densities of states, different, \cite{Moroz:1999} but this effect is expected to be small and ignored here, as well as in Ref.~\onlinecite{Liu:2014}.]

\vspace*{0.1in}

\subsection{Pairing Amplitude in 2D} \label{2D}
In 2D, all four components of the pairing amplitude are nonzero. While the explicit analytic expressions are too cumbersome to be included here, we find, again taking $\omega,\Delta\ll E_{so},E_{FR(S)}$, that the coefficients of Eq.~(\ref{coeffs}) are interrelated as follows:
\begin{equation} \label{coeffrels}
\begin{gathered}
c_{14,1}=-c_{13,1},\hspace*{0.1in}c_{16,2}=-c_{15,2},\hspace*{0.2in} c_{14,1},c_{16,2}\in\mathbb{R}, \\
c_{14,2}=-c_{13,2}^*=-c_{15,1}=c_{16,1}^*,\hspace*{0.52in} c_{14,2}\in\mathbb{C}.
\end{gathered}
\end{equation}
Given these relations, the matrices in Eq.~(\ref{amplitude}) can be expressed as
\vfill
\begin{widetext}
\begin{subequations} \label{matrices}
\begin{align}
f^0&=\left(\begin{array}{cc} c_{14,1} & \text{Re}[c_{14,2}] \\ \text{Re}[c_{14,2}] & c_{16,2} \end{array}\right), \\
f^x&=\left(\begin{array}{cc} -\sin\theta_1c_{14,1} & -\text{Im}[\sgn\omega e^{i\theta_1\,\sgn\omega}c_{14,2}] \\ -\text{Im}[\sgn\omega e^{-i\theta_2\,\sgn\omega}c_{14,2}] & \sin\theta_2c_{16,2} \end{array}\right), \\ 
f^y&=\left(\begin{array}{cc} \cos\theta_1c_{14,1} & \text{Re}[e^{i\theta_1\,\sgn\omega}c_{14,2}] \\ -\text{Re}[e^{-i\theta_2\,\sgn\omega}c_{14,2}] & -\cos\theta_2c_{16,2} \end{array}\right)\sgn\omega, \\
f^z&=\left(\begin{array}{cc} 0 & i\,\text{Im}[c_{14,2}] \\ -i\,\text{Im}[c_{14,2}] & 0 \end{array}\right),
\end{align}
\end{subequations}
\end{widetext} 
where we define $\theta_{1(2)}=\sin^{-1}(k_y/k_{F1(2)})$. We see that $f^z\propto\sigma_y$, while the remaining components contain both diagonal and off-diagonal terms. We therefore conclude that intraband Andreev reflection processes contribute to the singlet component and to only those triplet components that have $\bo{d}\parallel\bo{g}$, while interband processes contribute to all types of pairing. Additionally, each of the pairing components has a definite symmetry with respect to $k_y$; $s$ and $d_y$ are even functions of $k_y$, while $d_x$ and $d_z$ are odd. So, while a single trajectory parameterized by $k_y$ can produce all four pairing components, only the components $s$ and $d_y$ are nonzero when averaged over all possible trajectories.

To facilitate a quantitative comparison of the induced amplitudes of different symmetry in 2D junctions, we define a dimensionless analog to the angular-averaged quasiclassical Green's function, whereby we integrate the Gor'kov Green's function over the momentum,
\begin{equation} \label{def}
\hat \F^{\dagger}_\omega(x)=\frac{1}{m_R}\int_{-\infty}^\infty\frac{dk_y}{2\pi}\hat{\mathcal{F}}_{\omega,k_y}^{\dagger}(x,x).
\end{equation}
Note that we explicitly integrate over $k_y$, while the integration over $k_x$ is done implicitly by setting $x=x'$ in the nonlocal solution given in Eq.~(\ref{Fsol2}). In addition to allowing a quantitative comparison between singlet and triplet components, integrating over the momentum has the added benefit of picking out the odd-frequency triplet terms; this follows directly from the Pauli principle, as any triplet pairing amplitude is a sum of an even-frequency, odd-momentum component and an odd-frequency, even-momentum component.

We calculate the Green's function defined in Eq.~(\ref{def}) numerically without making any approximations. The pairing amplitude at a given $\omega$ is determined by four parameters: $m_R/m_S$, $E_{FR}/E_{FS}$, $E_{so}/E_{FR}$, and $E_{FS}/\Delta$. Figures~\ref{comparison}(a) and \ref{comparison}(b) illustrate the effects of SOC on the induced singlet and triplet amplitudes, respectively. While the magnitude of the pairing amplitude is largely determined by the Andreev reflection coefficient, which is controlled by all four parameters, we find that $E_{so}/E_{FR}$ is the only parameter that controls the singlet-to-triplet ratio. Figure~\ref{comparison}(c) compares the singlet and triplet pairing amplitudes using parameters appropriate for a giant Rashba semiconductor BiTeI, where $E_{so}/E_{FR}\approx2$.\cite{VanGennep:2014} As is seen from the plot, the singlet and triplet amplitudes can be comparable in magnitude in a real physical system. 

\begin{figure}[t!]
\includegraphics[width=\linewidth]{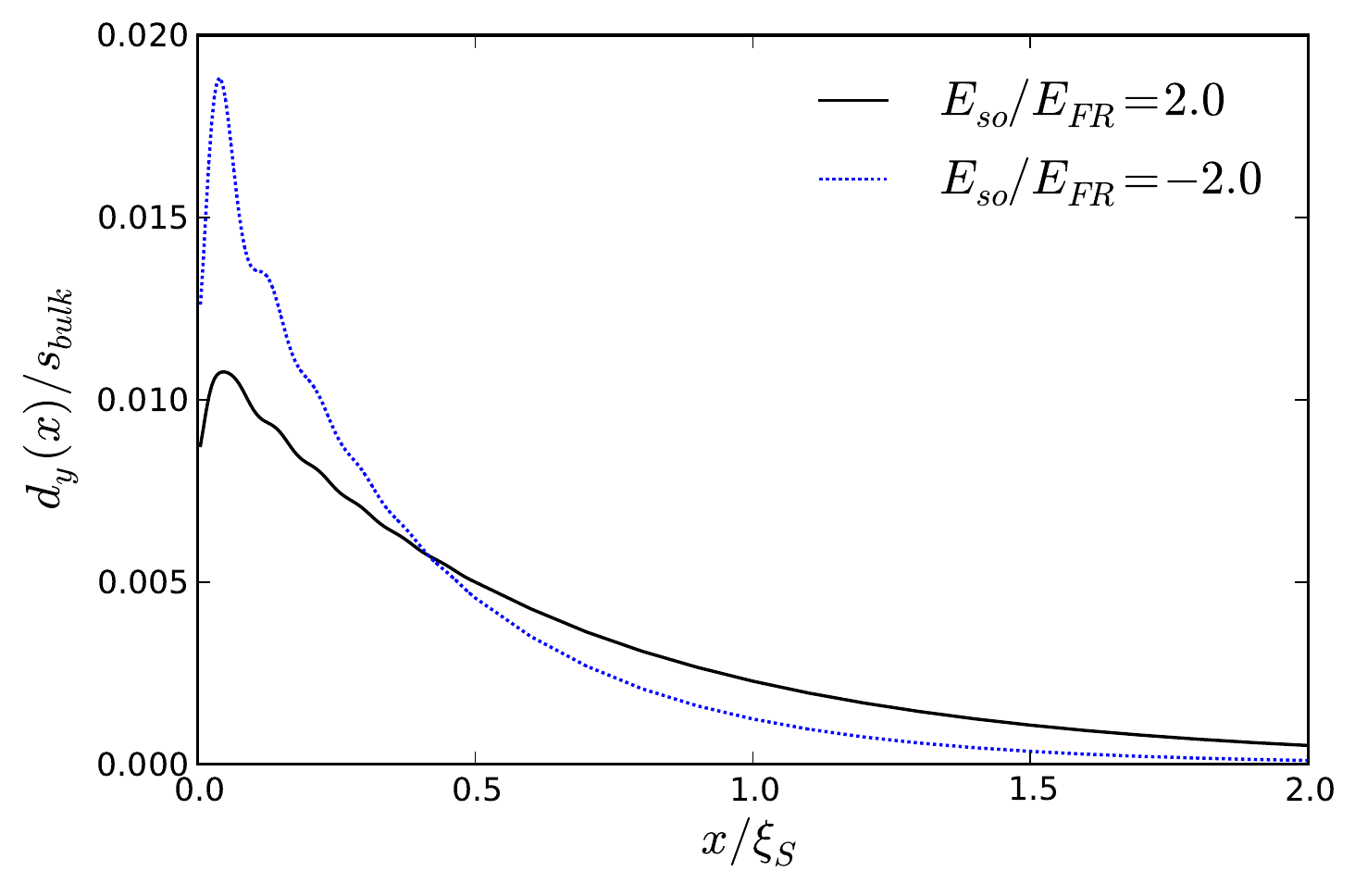}
\caption{\label{comparison2} (Color online) Quantitative comparison between induced triplet pairing amplitude in two-subband (solid line) and one-subband (dotted line) cases. Chosen parameters are the same as those in Fig.~\ref{comparison}, with only a change in the sign of $E_{FR}/E_{FS}$ for the one-subband case.
}
\end{figure}

As discussed previously in Sec.~\ref{S/R}, the generation of a triplet pairing component does not require occupation of both Rashba subbands. Figure~\ref{comparison2} shows that the induced triplet component is qualitatively similar when only a single subband is occupied. This observation is also relevant for BiTeI, as, in samples studied in Ref.~\onlinecite{VanGennep:2014}, only the lowest Rashba subband is occupied at ambient pressure.

\subsection{Recovering the Quasiclassical Limit} \label{quasiclassics}
We will now show that the induced triplet pairing amplitude vanishes to leading order in the quasiclassical limit, where SOC is taken to be weak ($\lambda\ll v_F$) and the splitting of the Fermi surfaces is neglected. If there is no Fermi surface mismatch between superconductor and Rashba metal ($m_R=m_S$ and $E_{FR}=E_{FS}$), the system can be described by a single Fermi momentum $k_F$ and a single angle $\theta=\sin^{-1}(k_y/k_F)$. Expanding the coefficients from Eq.~(\ref{coeffrels}) to linear order in $\lambda/v_F$, we find
\begin{equation}
\begin{gathered}
c_{14,1}=c_{16,2}=-\frac{\Delta}{2v_F\cos\theta(\Omega+|\omega|)}+\mathcal{O}(\lambda^2/v_F^2), \\
c_{14,2}=\mathcal{O}(\lambda^2/v_F^2),
\end{gathered}
\end{equation}
Because there is only one Fermi momentum, we can rewrite Eq.~(\ref{amplitude}) as
\begin{equation}
(s,\bo{d})=e^{-ik^*sx}e^{iksx'}\sum_{\alpha,\beta=1}^2\bigl[f^0_{\alpha\beta}(k_y,\omega),\bo{f}_{\alpha\beta}(k_y,\omega)\bigr], \\
\end{equation}
where $k^2=2m(i\omega+E_F)-k_y^2$ and $s=\sgn\omega$. We immediately see from Eqs.~(\ref{matrices}) that all three triplet components vanish to first order in $\lambda/v_F$. The nonzero triplet pairing amplitude occurs to order $\lambda^2/v_F^2$ (see also Refs. \onlinecite{Bergeret:2015,Konschelle:2015}) and is very small in materials with weak SOC. Therefore, in order to achieve a sizable singlet-triplet mixing, one needs to study the S/R proximity effect beyond the quasiclassical limit.

\section{Conclusions} \label{Conclusions}
We provided a qualitative physical argument, based on the consideration of possible Andreev reflection processes, for the presence of  triplet Cooper pairing in a Rashba material placed in proximity to a conventional superconductor. We also proved the existence of this triplet state by solving the exact Gor'kov equations. Even though the triplet state vanishes in 1D when integrated over the momentum, it can be comparable in magnitude to the singlet component in 2D when $E_{so}\sim E_{FR}$.

Because triplet pairing occurs only to order $\lambda^2/v_F^2$ in the quasiclassical limit, where $\lambda$ is the spin-orbit coupling constant and $v_F$ is the Fermi velocity, this work is most relevant to materials with large $\lambda$. For example, the spin-orbit constant in the bulk Rashba semiconductor BiTeI has been reported as $\lambda\sim1$ eV$\cdot$\AA, corresponding to a spin-orbit energy of $E_{so}=m^*\lambda^2/2\hbar^2\sim100$ meV ($m^*\sim0.1m_e$). \cite{VanGennep:2014} More conventional semiconductors provide a wide range of spin-orbit coupling strengths. While GaAs/AlGaAs quantum wells are weak spin-orbit materials with $\lambda\sim1$ meV$\cdot$\AA,{\cite{Miller:2003}} other semiconductor heterostructures, such as InAs/InAlAs \cite{Grundler:2000} and InSb/InAlSb \cite{Khodaparast:2004,Gilbertson:2009,Leontiadou:2011} quantum wells, have been reported to have $\lambda\sim0.1$ eV$\cdot$\AA. Additionally, Ge/Si core/shell, \cite{Hao:2010} InAs, \cite{Liang:2012} and InSb \cite{vanWeperen:2015} nanowires can reach spin-orbit strengths of $\lambda\sim0.1-1$ eV$\cdot$\AA. With so many materials belonging to the strong spin-orbit coupling regime, it is imperative to understand the proximity effect in such materials beyond the quasiclassical approximation.

Even though we do not treat disorder here explicitly, we can comment on its effect. In the presence of impurity scattering, the triplet Cooper pairs in a spin-orbit system are subjected to spin relaxation. Therefore, in the diffusive limit, where $\ell\ll\xi_S$ and $\ell\ll\ell_{so}$ ($\ell$ is the mean free path and $\ell_{so}\sim\hbar^2/m^*\lambda$ is the spin-orbit length), the decay length of the odd-frequency triplet component is $\sim\sqrt{D/\tau_{so}}$, where $\tau_{so}$ is the spin relaxation time and $D$ is the diffusion coefficient. In the diffusive limit of the Dyakonov-Perel mechanism, \cite{Dyakonov:1972} the spin relaxation time is given by $\tau_{so}\sim D/\ell_{so}^2$, so the wave function of the triplet Cooper pairs is expected to decay on the scale set by $\ell_{so}$. However, in materials with strong spin-orbit coupling for which the induced triplet pairing is relevant, the spin-orbit length is comparable to the Fermi wavelength; therefore, as long as these materials remain ``good metals" ($k_F\ell\gg1$), the results of this work should apply. Recent experiments have measured an electron mobility exceeding $200\,000$ cm$^2$/V$\cdot$s in an InSb/InAlSb quantum well, \cite{Yi:2015} thus demonstrating an ability to fabricate ultraclean materials with strong spin-orbit coupling.

\acknowledgments
We thank A. Kumar, S. Maiti, E. Rashba, and V. Zyuzin for helpful discussions and S. Bergeret, F. Konschelle, and I. Tokatly for bringing Refs.~\onlinecite{Bergeret:2015,Konschelle:2015} to our attention. This work was supported by the National Science Foundation via grant NSF DMR-1308972. DLM is the Ulam Scholar at the Center for Nonlinear Studies, Los Alamos National Laboratory.

\begin{widetext}
\appendix
\section{Triplet pairing amplitude as a consequence of lifted spin degeneracy} \label{appendix}
In this appendix, we prove that a nonzero triplet pairing amplitude is induced in any 1D proximity junction with lifted spin degeneracy. For our purposes here, we assume that spin remains an eigenstate of the Hamiltonian, as is the case for both a ferromagnet and a 1D Rashba material. By calculating the BdG wave functions describing the Andreev reflection of an electron to a hole, we will be able to determine the pairing amplitude and show that the triplet component is always nonzero when the spin degeneracy is lifted.

In any nonsuperconducting material where spin is an eigenstate of the Hamiltonian, the scattering wave function describing an electron incident on the superconducting interface can be expressed as
\begin{equation} \label{psiN}
\psi_N(x,\varepsilon)=\psi_i(x,\varepsilon)+a_\uparrow(\varepsilon)\left(\begin{array}{c} 0 \\ 0 \\ 1 \\ 0 \end{array}\right)e^{ik_{\uparrow h}^Lx}+a_\downarrow(\varepsilon)\left(\begin{array}{c} 0 \\ 0 \\ 0 \\ 1 \end{array}\right)e^{ik_{\downarrow h}^Lx}+r_\uparrow(\varepsilon)\left(\begin{array}{c} 1 \\ 0 \\ 0 \\ 0 \end{array}\right)e^{ik_{\uparrow e}^Lx}+r_\downarrow(\varepsilon)\left(\begin{array}{c} 0 \\ 1 \\ 0 \\ 0 \end{array}\right)e^{ik_{\downarrow e}^Lx},
\end{equation}
where $\psi_i(x,\varepsilon)$ is the wave function of the incident electron, $k_{\uparrow(\downarrow)e(h)}^{R(L)}$ is the momentum of a right-moving (left-moving) spin-up (spin-down) electron (hole), $a_{\uparrow(\downarrow)}(\varepsilon)$ are the Andreev reflection amplitudes, and $r_{\uparrow(\downarrow)}(\varepsilon)$ are the normal reflection amplitudes. On the superconducting side, the wave function is given by
\begin{equation}
\psi_S(x,\varepsilon)=t_1(\varepsilon)\left(\begin{array}{c} u_0 \\ 0 \\ 0 \\ v_0 \end{array}\right)e^{iq_+x}+t_2(\varepsilon)\left(\begin{array}{c} 0 \\ u_0 \\ -v_0 \\ 0 \end{array}\right)e^{iq_+x}+t_3(\varepsilon)\left(\begin{array}{c} v_0 \\ 0 \\ 0 \\ u_0 \end{array}\right)e^{-iq_-x}+t_4(\varepsilon)\left(\begin{array}{c} 0 \\ -v_0 \\ u_0 \\ 0 \end{array}\right)e^{-iq_-x},
\end{equation}
where $q^2_\pm=2m_S(E_{FS}\pm i\sqrt{\Delta^2-\varepsilon^2})$, $u_0^2=(1+i\sqrt{\Delta^2-\varepsilon^2}/\varepsilon)/2$ and $v_0^2=(1-i\sqrt{\Delta^2-\varepsilon^2}/\varepsilon)/2$ are the usual BCS coherence factors, and the $t_i(\varepsilon)$ are transmission amplitudes. Boundary conditions to be imposed on the wave functions are
\begin{subequations}
\begin{gather}
\psi_N(0)=\psi_S(0), \\
\hat v_N\psi_N(0)=\hat v_S\psi_S(0).
\end{gather}
\end{subequations}
The velocity operator is a $4\times4$ matrix given by $\hat{v}=\partial H/\partial(-i\partial_x)$, where $H$ is the Hamiltonian of the BdG equation [$H\psi(x,\varepsilon)=\varepsilon\psi(x,\varepsilon)$].

We first consider the scattering of an incident spin-up electron. In that case, $[\psi_i(x,\varepsilon)]^T=(1,0,0,0)e^{ik_{\uparrow e}^Rx}$. Imposing the boundary conditions gives us a set of four equations ($a_\uparrow=r_\downarrow=0$ for this choice of incident wave function):
\begin{equation} \label{boundary1}
\begin{aligned}
1+r_\uparrow&=t_1u_0+t_3v_0, \hspace*{0.5in} v_{F\uparrow}(1-r_\uparrow)=v_{FS}(t_1u_0-t_3v_0), \\
a_\downarrow&=t_1v_0+t_3u_0,\hspace*{0.84in} v_{F\downarrow}a_\downarrow=v_{FS}(t_1v_0-t_3u_0),
\end{aligned}
\end{equation}
where $v_{F\uparrow(\downarrow)}$ is the Fermi velocity of the spin-up (spin-down) band of the nonsuperconducting material and $v_{FS}$ is the Fermi velocity of the superconductor. In obtaining Eq.~(\ref{boundary1}), we expanded in the limit $\varepsilon,\Delta\ll E_{FN(S)}$. We also must consider the scattering of an incident spin-down electron, with wave function given by $[\psi_i(x,\varepsilon)]^T=(0,1,0,0)e^{ik_{\uparrow e}^Rx}$. The boundary conditions for this second case are
\begin{equation} \label{boundary2}
\begin{aligned}
1+r_\downarrow&=t_2u_0-t_4v_0, \hspace*{0.5in} v_{F\downarrow}(1-r_\downarrow)=v_{FS}(t_2u_0+t_4v_0), \\
a_\uparrow&=-t_2v_0+t_4u_0,\hspace*{0.63in} -v_{F\uparrow}a_\uparrow=v_{FS}(t_2v_0+t_4u_0).
\end{aligned}
\end{equation}
By solving Eqs.~(\ref{boundary1}) and (\ref{boundary2}), we find that the Andreev and normal reflection amplitudes are related by
\begin{subequations} \label{relations}
\begin{gather}
a_\downarrow(\varepsilon)=-\frac{v_{F\uparrow}}{v_{F\downarrow}}a_\uparrow(\varepsilon), \\
r_\uparrow(\varepsilon)=\frac{(u_0^2-v_0^2)(v_{F\uparrow}v_{F\downarrow}-v_{FS}^2)+v_{FS}(v_{F\uparrow}-v_{F\downarrow})}{(u_0^2-v_0^2)(v_{F\uparrow}v_{F\downarrow}-v_{FS}^2)-v_{FS}(v_{F\uparrow}-v_{F\downarrow})}r_\downarrow(\varepsilon).
\end{gather}
\end{subequations}
The relations of Eq.~(\ref{relations}) will allow us to prove the existence of a triplet pairing component.

The pairing amplitude is given in terms of the BdG wave functions $u_\alpha(x,\varepsilon)$ (electrons) and $v_\alpha(x,\varepsilon)$ (holes) by
\begin{equation}
\F^\dagger_{\alpha\beta}(x,x',\varepsilon)\sim v_\alpha(x,\varepsilon)u_\beta^*(x',\varepsilon).
\end{equation}
Therefore, the triplet component of the pairing amplitude with zero spin projection is
\begin{equation}
\F^\dagger_t(x,x',\varepsilon)=\frac{1}{2}\left[\F_\ud^\dagger(x,x',\varepsilon)+\F_\du^\dagger(x,x',\varepsilon)\right]\sim v_\uparrow(x,\varepsilon)u^*_\downarrow(x',\varepsilon)+v_\downarrow(x,\varepsilon)u^*_\uparrow(x',\varepsilon).
\end{equation}
These two terms correspond to the two scattering processes considered above. Substituting the wave functions from Eq.~(\ref{psiN}), we find that
\begin{equation} \label{exp1}
\F^\dagger_t(x,x',\varepsilon)\sim a_\uparrow(\varepsilon)\left[e^{-ik_{\downarrow e}^Rx'}+r_\downarrow^*(\varepsilon)e^{-ik_{\downarrow e}^Lx'}\right]e^{ik_{\uparrow h}^Lx}+a_\downarrow(\varepsilon)\left[e^{-ik_{\uparrow e}^Rx'}+r_\uparrow^*(\varepsilon)e^{-ik_{\uparrow e}^Lx'}\right]e^{ik_{\downarrow h}^Lx}.
\end{equation}
Given the relations in Eq.~(\ref{relations}), we can reexpress Eq.~(\ref{exp1}) as
\begin{equation} \label{exp2}
\begin{aligned}
\F^\dagger_t(x,x',\varepsilon)&\sim a_\uparrow(\varepsilon)\left(e^{ik_{\uparrow h}^Lx}e^{-ik_{\downarrow e}^Rx'}-\frac{v_{F\uparrow}}{v_{F\downarrow}}e^{ik_{\downarrow h}^Lx}e^{-ik_{\uparrow e}^Rx'}\right) \\
	&+a_\uparrow(\varepsilon)r_\downarrow^*(\varepsilon)\left(e^{ik_{\uparrow h}^Lx}e^{-ik_{\downarrow e}^Lx'}-\frac{v_{F\uparrow}}{v_{F\downarrow}}\frac{\sgn{\varepsilon-\Delta}(u_0^2-v_0^2)(v_{F\uparrow}v_{F\downarrow}-v_{FS}^2)+v_{FS}(v_{F\uparrow}-v_{F\downarrow})}{\sgn{\varepsilon-\Delta}(u_0^2-v_0^2)(v_{F\uparrow}v_{F\downarrow}-v_{FS}^2)-v_{FS}(v_{F\uparrow}-v_{F\downarrow})}e^{ik_{\downarrow h}^Lx}e^{-ik_{\uparrow e}^Lx'}\right).
\end{aligned}
\end{equation}
Note that the triplet amplitude only vanishes when the two spin bands are degenerate, $k_\uparrow=k_\downarrow$ and $v_{F\uparrow}=v_{F\downarrow}$. Therefore, triplet pairing is induced by the proximity effect any time the spin degeneracy is lifted.

\end{widetext}

\bibliography{bibRashbaTriplet}

\begin{thebibliography}{50}%
\makeatletter
\providecommand \@ifxundefined [1]{%
 \@ifx{#1\undefined}
}%
\providecommand \@ifnum [1]{%
 \ifnum #1\expandafter \@firstoftwo
 \else \expandafter \@secondoftwo
 \fi
}%
\providecommand \@ifx [1]{%
 \ifx #1\expandafter \@firstoftwo
 \else \expandafter \@secondoftwo
 \fi
}%
\providecommand \natexlab [1]{#1}%
\providecommand \enquote  [1]{``#1''}%
\providecommand \bibnamefont  [1]{#1}%
\providecommand \bibfnamefont [1]{#1}%
\providecommand \citenamefont [1]{#1}%
\providecommand \href@noop [0]{\@secondoftwo}%
\providecommand \href [0]{\begingroup \@sanitize@url \@href}%
\providecommand \@href[1]{\@@startlink{#1}\@@href}%
\providecommand \@@href[1]{\endgroup#1\@@endlink}%
\providecommand \@sanitize@url [0]{\catcode `\\12\catcode `\$12\catcode
  `\&12\catcode `\#12\catcode `\^12\catcode `\_12\catcode `\%12\relax}%
\providecommand \@@startlink[1]{}%
\providecommand \@@endlink[0]{}%
\providecommand \url  [0]{\begingroup\@sanitize@url \@url }%
\providecommand \@url [1]{\endgroup\@href {#1}{\urlprefix }}%
\providecommand \urlprefix  [0]{URL }%
\providecommand \Eprint [0]{\href }%
\providecommand \doibase [0]{http://dx.doi.org/}%
\providecommand \selectlanguage [0]{\@gobble}%
\providecommand \bibinfo  [0]{\@secondoftwo}%
\providecommand \bibfield  [0]{\@secondoftwo}%
\providecommand \translation [1]{[#1]}%
\providecommand \BibitemOpen [0]{}%
\providecommand \bibitemStop [0]{}%
\providecommand \bibitemNoStop [0]{.\EOS\space}%
\providecommand \EOS [0]{\spacefactor3000\relax}%
\providecommand \BibitemShut  [1]{\csname bibitem#1\endcsname}%
\let\auto@bib@innerbib\@empty
\bibitem [{\citenamefont {Eschrig}(2011)}]{Eschrig:2011}%
  \BibitemOpen
  \bibfield  {author} {\bibinfo {author} {\bibfnamefont {M.}~\bibnamefont
  {Eschrig}},\ }\href@noop {} {\bibfield  {journal} {\bibinfo  {journal}
  {Physics Today}\ }\textbf {\bibinfo {volume} {{\bf 64}}},\ \bibinfo {pages}
  {43} (\bibinfo {year} {2011})}\BibitemShut {NoStop}%
\bibitem [{\citenamefont {Klose}\ \emph {et~al.}(2012)\citenamefont {Klose},
  \citenamefont {Khaire}, \citenamefont {Wang}, \citenamefont {Pratt},
  \citenamefont {Birge}, \citenamefont {McMorran}, \citenamefont {Ginley},
  \citenamefont {Borchers}, \citenamefont {Kirby}, \citenamefont {Maranville},\
  and\ \citenamefont {Unguris}}]{Klose:2012}%
  \BibitemOpen
  \bibfield  {author} {\bibinfo {author} {\bibfnamefont {C.}~\bibnamefont
  {Klose}}, \bibinfo {author} {\bibfnamefont {T.~S.}\ \bibnamefont {Khaire}},
  \bibinfo {author} {\bibfnamefont {Y.}~\bibnamefont {Wang}}, \bibinfo {author}
  {\bibfnamefont {W.~P.}\ \bibnamefont {Pratt}}, \bibinfo {author}
  {\bibfnamefont {N.~O.}\ \bibnamefont {Birge}}, \bibinfo {author}
  {\bibfnamefont {B.~J.}\ \bibnamefont {McMorran}}, \bibinfo {author}
  {\bibfnamefont {T.~P.}\ \bibnamefont {Ginley}}, \bibinfo {author}
  {\bibfnamefont {J.~A.}\ \bibnamefont {Borchers}}, \bibinfo {author}
  {\bibfnamefont {B.~J.}\ \bibnamefont {Kirby}}, \bibinfo {author}
  {\bibfnamefont {B.~B.}\ \bibnamefont {Maranville}}, \ and\ \bibinfo {author}
  {\bibfnamefont {J.}~\bibnamefont {Unguris}},\ }\href {\doibase
  10.1103/PhysRevLett.108.127002} {\bibfield  {journal} {\bibinfo  {journal}
  {Phys. Rev. Lett.}\ }\textbf {\bibinfo {volume} {{\bf 108}}},\ \bibinfo
  {pages} {127002} (\bibinfo {year} {2012})}\BibitemShut {NoStop}%
\bibitem [{\citenamefont {Quay}\ \emph {et~al.}(2013)\citenamefont {Quay},
  \citenamefont {Chevallier}, \citenamefont {Bena},\ and\ \citenamefont
  {Aprili}}]{Quay:2013}%
  \BibitemOpen
  \bibfield  {author} {\bibinfo {author} {\bibfnamefont {C.~H.~L.}\
  \bibnamefont {Quay}}, \bibinfo {author} {\bibfnamefont {D.}~\bibnamefont
  {Chevallier}}, \bibinfo {author} {\bibfnamefont {C.}~\bibnamefont {Bena}}, \
  and\ \bibinfo {author} {\bibfnamefont {M.}~\bibnamefont {Aprili}},\ }\href
  {http://dx.doi.org/10.1038/nphys2518} {\bibfield  {journal} {\bibinfo
  {journal} {Nat. Phys.}\ }\textbf {\bibinfo {volume} {{\bf 9}}},\ \bibinfo
  {pages} {84} (\bibinfo {year} {2013})}\BibitemShut {NoStop}%
\bibitem [{\citenamefont {Banerjee}\ \emph
  {et~al.}(2014{\natexlab{a}})\citenamefont {Banerjee}, \citenamefont {Smiet},
  \citenamefont {Smits}, \citenamefont {Ozaeta}, \citenamefont {Bergeret},
  \citenamefont {Blamire},\ and\ \citenamefont {Robinson}}]{Banerjee:2014}%
  \BibitemOpen
  \bibfield  {author} {\bibinfo {author} {\bibfnamefont {N.}~\bibnamefont
  {Banerjee}}, \bibinfo {author} {\bibfnamefont {C.~B.}\ \bibnamefont {Smiet}},
  \bibinfo {author} {\bibfnamefont {R.~G.~J.}\ \bibnamefont {Smits}}, \bibinfo
  {author} {\bibfnamefont {A.}~\bibnamefont {Ozaeta}}, \bibinfo {author}
  {\bibfnamefont {F.~S.}\ \bibnamefont {Bergeret}}, \bibinfo {author}
  {\bibfnamefont {M.~G.}\ \bibnamefont {Blamire}}, \ and\ \bibinfo {author}
  {\bibfnamefont {J.~W.~A.}\ \bibnamefont {Robinson}},\ }\href
  {http://dx.doi.org/10.1038/ncomms4048} {\bibfield  {journal} {\bibinfo
  {journal} {Nat. Commun.}\ }\textbf {\bibinfo {volume} {{\bf 5}}},\ \bibinfo
  {pages} {3048} (\bibinfo {year} {2014}{\natexlab{a}})}\BibitemShut {NoStop}%
\bibitem [{\citenamefont {Banerjee}\ \emph
  {et~al.}(2014{\natexlab{b}})\citenamefont {Banerjee}, \citenamefont
  {Robinson},\ and\ \citenamefont {Blamire}}]{Banerjee:2014ly}%
  \BibitemOpen
  \bibfield  {author} {\bibinfo {author} {\bibfnamefont {N.}~\bibnamefont
  {Banerjee}}, \bibinfo {author} {\bibfnamefont {J.~W.~A.}\ \bibnamefont
  {Robinson}}, \ and\ \bibinfo {author} {\bibfnamefont {M.~G.}\ \bibnamefont
  {Blamire}},\ }\href {http://dx.doi.org/10.1038/ncomms5771} {\bibfield
  {journal} {\bibinfo  {journal} {Nat. Commun.}\ }\textbf {\bibinfo {volume}
  {{\bf 5}}},\ \bibinfo {pages} {4771} (\bibinfo {year}
  {2014}{\natexlab{b}})}\BibitemShut {NoStop}%
\bibitem [{\citenamefont {Linder}\ and\ \citenamefont
  {Robinson}(2015)}]{Linder:2015}%
  \BibitemOpen
  \bibfield  {author} {\bibinfo {author} {\bibfnamefont {J.}~\bibnamefont
  {Linder}}\ and\ \bibinfo {author} {\bibfnamefont {J.~W.~A.}\ \bibnamefont
  {Robinson}},\ }\href {http://dx.doi.org/10.1038/nphys3242} {\bibfield
  {journal} {\bibinfo  {journal} {Nat. Phys.}\ }\textbf {\bibinfo {volume}
  {{\bf 11}}},\ \bibinfo {pages} {307} (\bibinfo {year} {2015})}\BibitemShut
  {NoStop}%
\bibitem [{\citenamefont {Datta}\ and\ \citenamefont {Das}(1990)}]{Datta:1990}%
  \BibitemOpen
  \bibfield  {author} {\bibinfo {author} {\bibfnamefont {S.}~\bibnamefont
  {Datta}}\ and\ \bibinfo {author} {\bibfnamefont {B.}~\bibnamefont {Das}},\
  }\href {\doibase http://dx.doi.org/10.1063/1.102730} {\bibfield  {journal}
  {\bibinfo  {journal} {Appl. Phys. Lett.}\ }\textbf {\bibinfo {volume} {{\bf
  56}}},\ \bibinfo {pages} {665} (\bibinfo {year} {1990})}\BibitemShut
  {NoStop}%
\bibitem [{\citenamefont {\ifmmode \check{Z}\else
  \v{Z}\fi{}uti\ifmmode~\acute{c}\else \'{c}\fi{}}\ \emph
  {et~al.}(2004)\citenamefont {\ifmmode \check{Z}\else
  \v{Z}\fi{}uti\ifmmode~\acute{c}\else \'{c}\fi{}}, \citenamefont {Fabian},\
  and\ \citenamefont {Das~Sarma}}]{Zutic:2004}%
  \BibitemOpen
  \bibfield  {author} {\bibinfo {author} {\bibfnamefont {I.}~\bibnamefont
  {\ifmmode \check{Z}\else \v{Z}\fi{}uti\ifmmode~\acute{c}\else \'{c}\fi{}}},
  \bibinfo {author} {\bibfnamefont {J.}~\bibnamefont {Fabian}}, \ and\ \bibinfo
  {author} {\bibfnamefont {S.}~\bibnamefont {Das~Sarma}},\ }\href {\doibase
  10.1103/RevModPhys.76.323} {\bibfield  {journal} {\bibinfo  {journal} {Rev.
  Mod. Phys.}\ }\textbf {\bibinfo {volume} {{\bf 76}}},\ \bibinfo {pages} {323}
  (\bibinfo {year} {2004})}\BibitemShut {NoStop}%
\bibitem [{\citenamefont {Buzdin}(2005)}]{Buzdin:2005}%
  \BibitemOpen
  \bibfield  {author} {\bibinfo {author} {\bibfnamefont {A.}~\bibnamefont
  {Buzdin}},\ }\href {\doibase 10.1103/RevModPhys.77.935} {\bibfield  {journal}
  {\bibinfo  {journal} {Rev. Mod. Phys.}\ }\textbf {\bibinfo {volume} {{\bf
  77}}},\ \bibinfo {pages} {935} (\bibinfo {year} {2005})}\BibitemShut
  {NoStop}%
\bibitem [{\citenamefont {Bergeret}\ \emph {et~al.}(2005)\citenamefont
  {Bergeret}, \citenamefont {Volkov},\ and\ \citenamefont
  {Efetov}}]{Bergeret:2005}%
  \BibitemOpen
  \bibfield  {author} {\bibinfo {author} {\bibfnamefont {F.}~\bibnamefont
  {Bergeret}}, \bibinfo {author} {\bibfnamefont {A.}~\bibnamefont {Volkov}}, \
  and\ \bibinfo {author} {\bibfnamefont {K.}~\bibnamefont {Efetov}},\ }\href
  {\doibase 10.1103/RevModPhys.77.1321} {\bibfield  {journal} {\bibinfo
  {journal} {Rev. Mod. Phys.}\ }\textbf {\bibinfo {volume} {{\bf 77}}},\
  \bibinfo {pages} {1321} (\bibinfo {year} {2005})}\BibitemShut {NoStop}%
\bibitem [{\citenamefont {Bergeret}\ \emph {et~al.}(2001)\citenamefont
  {Bergeret}, \citenamefont {Volkov},\ and\ \citenamefont
  {Efetov}}]{Bergeret:2001}%
  \BibitemOpen
  \bibfield  {author} {\bibinfo {author} {\bibfnamefont {F.~S.}\ \bibnamefont
  {Bergeret}}, \bibinfo {author} {\bibfnamefont {A.~F.}\ \bibnamefont
  {Volkov}}, \ and\ \bibinfo {author} {\bibfnamefont {K.~B.}\ \bibnamefont
  {Efetov}},\ }\href {\doibase 10.1103/PhysRevLett.86.4096} {\bibfield
  {journal} {\bibinfo  {journal} {Phys. Rev. Lett.}\ }\textbf {\bibinfo
  {volume} {{\bf 86}}},\ \bibinfo {pages} {4096} (\bibinfo {year}
  {2001})}\BibitemShut {NoStop}%
\bibitem [{\citenamefont {Volkov}\ \emph {et~al.}(2003)\citenamefont {Volkov},
  \citenamefont {Bergeret},\ and\ \citenamefont {Efetov}}]{Volkov:2003}%
  \BibitemOpen
  \bibfield  {author} {\bibinfo {author} {\bibfnamefont {A.~F.}\ \bibnamefont
  {Volkov}}, \bibinfo {author} {\bibfnamefont {F.~S.}\ \bibnamefont
  {Bergeret}}, \ and\ \bibinfo {author} {\bibfnamefont {K.~B.}\ \bibnamefont
  {Efetov}},\ }\href {\doibase 10.1103/PhysRevLett.90.117006} {\bibfield
  {journal} {\bibinfo  {journal} {Phys. Rev. Lett.}\ }\textbf {\bibinfo
  {volume} {{\bf 90}}},\ \bibinfo {pages} {117006} (\bibinfo {year}
  {2003})}\BibitemShut {NoStop}%
\bibitem [{\citenamefont {Yokoyama}\ \emph {et~al.}(2007)\citenamefont
  {Yokoyama}, \citenamefont {Tanaka},\ and\ \citenamefont
  {Golubov}}]{Yokoyama:2007}%
  \BibitemOpen
  \bibfield  {author} {\bibinfo {author} {\bibfnamefont {T.}~\bibnamefont
  {Yokoyama}}, \bibinfo {author} {\bibfnamefont {Y.}~\bibnamefont {Tanaka}}, \
  and\ \bibinfo {author} {\bibfnamefont {A.~A.}\ \bibnamefont {Golubov}},\
  }\href {\doibase 10.1103/PhysRevB.75.134510} {\bibfield  {journal} {\bibinfo
  {journal} {Phys. Rev. B}\ }\textbf {\bibinfo {volume} {{\bf 75}}},\ \bibinfo
  {pages} {134510} (\bibinfo {year} {2007})}\BibitemShut {NoStop}%
\bibitem [{\citenamefont {Edelstein}(1989)}]{Edelstein:1989}%
  \BibitemOpen
  \bibfield  {author} {\bibinfo {author} {\bibfnamefont {V.~M.}\ \bibnamefont
  {Edelstein}},\ }\href@noop {} {\bibfield  {journal} {\bibinfo  {journal} {Zh.
  Eksp. Teor. Fiz.}\ }\textbf {\bibinfo {volume} {{\bf 95}}},\ \bibinfo {pages}
  {2151} (\bibinfo {year} {1989})},\ \bibinfo {note} {[Sov. Phys. JETP {\bf
  68}, 1244 (1989)]}\BibitemShut {NoStop}%
\bibitem [{\citenamefont {Gor'kov}\ and\ \citenamefont
  {Rashba}(2001)}]{Gorkov:2001}%
  \BibitemOpen
  \bibfield  {author} {\bibinfo {author} {\bibfnamefont {L.~P.}\ \bibnamefont
  {Gor'kov}}\ and\ \bibinfo {author} {\bibfnamefont {E.~I.}\ \bibnamefont
  {Rashba}},\ }\href {\doibase 10.1103/PhysRevLett.87.037004} {\bibfield
  {journal} {\bibinfo  {journal} {Phys. Rev. Lett.}\ }\textbf {\bibinfo
  {volume} {{\bf 87}}},\ \bibinfo {pages} {037004} (\bibinfo {year}
  {2001})}\BibitemShut {NoStop}%
\bibitem [{\citenamefont {Frigeri}\ \emph {et~al.}(2004)\citenamefont
  {Frigeri}, \citenamefont {Agterberg}, \citenamefont {Koga},\ and\
  \citenamefont {Sigrist}}]{Frigeri:2004}%
  \BibitemOpen
  \bibfield  {author} {\bibinfo {author} {\bibfnamefont {P.~A.}\ \bibnamefont
  {Frigeri}}, \bibinfo {author} {\bibfnamefont {D.~F.}\ \bibnamefont
  {Agterberg}}, \bibinfo {author} {\bibfnamefont {A.}~\bibnamefont {Koga}}, \
  and\ \bibinfo {author} {\bibfnamefont {M.}~\bibnamefont {Sigrist}},\ }\href
  {\doibase 10.1103/PhysRevLett.92.097001} {\bibfield  {journal} {\bibinfo
  {journal} {Phys. Rev. Lett.}\ }\textbf {\bibinfo {volume} {{\bf 92}}},\
  \bibinfo {pages} {097001} (\bibinfo {year} {2004})}\BibitemShut {NoStop}%
\bibitem [{\citenamefont {Liu}\ \emph {et~al.}(2014)\citenamefont {Liu},
  \citenamefont {Jain},\ and\ \citenamefont {Liu}}]{Liu:2014}%
  \BibitemOpen
  \bibfield  {author} {\bibinfo {author} {\bibfnamefont {X.}~\bibnamefont
  {Liu}}, \bibinfo {author} {\bibfnamefont {J.~K.}\ \bibnamefont {Jain}}, \
  and\ \bibinfo {author} {\bibfnamefont {C.-X.}\ \bibnamefont {Liu}},\ }\href
  {\doibase 10.1103/PhysRevLett.113.227002} {\bibfield  {journal} {\bibinfo
  {journal} {Phys. Rev. Lett.}\ }\textbf {\bibinfo {volume} {{\bf 113}}},\
  \bibinfo {pages} {227002} (\bibinfo {year} {2014})}\BibitemShut {NoStop}%
\bibitem [{\citenamefont {Bychkov}\ and\ \citenamefont
  {Rashba}(1984)}]{Bychkov:1984}%
  \BibitemOpen
  \bibfield  {author} {\bibinfo {author} {\bibfnamefont {Y.~A.}\ \bibnamefont
  {Bychkov}}\ and\ \bibinfo {author} {\bibfnamefont {E.~I.}\ \bibnamefont
  {Rashba}},\ }\href {http://stacks.iop.org/0022-3719/17/i=33/a=015} {\bibfield
   {journal} {\bibinfo  {journal} {J. Phys. C: Solid State Phys.}\ }\textbf
  {\bibinfo {volume} {{\bf 17}}},\ \bibinfo {pages} {6039} (\bibinfo {year}
  {1984})}\BibitemShut {NoStop}%
\bibitem [{\citenamefont {Bergeret}\ and\ \citenamefont
  {Tokatly}(2013)}]{Bergeret:2013}%
  \BibitemOpen
  \bibfield  {author} {\bibinfo {author} {\bibfnamefont {F.~S.}\ \bibnamefont
  {Bergeret}}\ and\ \bibinfo {author} {\bibfnamefont {I.~V.}\ \bibnamefont
  {Tokatly}},\ }\href {\doibase 10.1103/PhysRevLett.110.117003} {\bibfield
  {journal} {\bibinfo  {journal} {Phys. Rev. Lett.}\ }\textbf {\bibinfo
  {volume} {{\bf 110}}},\ \bibinfo {pages} {117003} (\bibinfo {year}
  {2013})}\BibitemShut {NoStop}%
\bibitem [{\citenamefont {Bergeret}\ and\ \citenamefont
  {Tokatly}(2014)}]{Bergeret:2014}%
  \BibitemOpen
  \bibfield  {author} {\bibinfo {author} {\bibfnamefont {F.~S.}\ \bibnamefont
  {Bergeret}}\ and\ \bibinfo {author} {\bibfnamefont {I.~V.}\ \bibnamefont
  {Tokatly}},\ }\href {\doibase 10.1103/PhysRevB.89.134517} {\bibfield
  {journal} {\bibinfo  {journal} {Phys. Rev. B}\ }\textbf {\bibinfo {volume}
  {{\bf 89}}},\ \bibinfo {pages} {134517} (\bibinfo {year} {2014})}\BibitemShut
  {NoStop}%
\bibitem [{Ber()}]{BergeretNote}%
  \BibitemOpen
  \href@noop {} {}\bibinfo {note} {This can be seen if the ferromagnetic
  exchange field is eliminated.}\BibitemShut {Stop}%
\bibitem [{\citenamefont {Bergeret}\ and\ \citenamefont
  {Tokatly}(2015)}]{Bergeret:2015}%
  \BibitemOpen
  \bibfield  {author} {\bibinfo {author} {\bibfnamefont {F.~S.}\ \bibnamefont
  {Bergeret}}\ and\ \bibinfo {author} {\bibfnamefont {I.~V.}\ \bibnamefont
  {Tokatly}},\ }\href {http://stacks.iop.org/0295-5075/110/i=5/a=57005}
  {\bibfield  {journal} {\bibinfo  {journal} {Europhys. Lett.}\ }\textbf
  {\bibinfo {volume} {{\bf 110}}},\ \bibinfo {pages} {57005} (\bibinfo {year}
  {2015})}\BibitemShut {NoStop}%
\bibitem [{\citenamefont {Konschelle}\ \emph {et~al.}(2015)\citenamefont
  {Konschelle}, \citenamefont {Tokatly},\ and\ \citenamefont
  {Bergeret}}]{Konschelle:2015}%
  \BibitemOpen
  \bibfield  {author} {\bibinfo {author} {\bibfnamefont {F.}~\bibnamefont
  {Konschelle}}, \bibinfo {author} {\bibfnamefont {I.~V.}\ \bibnamefont
  {Tokatly}}, \ and\ \bibinfo {author} {\bibfnamefont {F.~S.}\ \bibnamefont
  {Bergeret}},\ }\href {\doibase 10.1103/PhysRevB.92.125443} {\bibfield
  {journal} {\bibinfo  {journal} {Phys. Rev. B}\ }\textbf {\bibinfo {volume}
  {{\bf 92}}},\ \bibinfo {pages} {125443} (\bibinfo {year} {2015})}\BibitemShut
  {NoStop}%
\bibitem [{\citenamefont {Edelstein}(2003)}]{Edelstein:2003}%
  \BibitemOpen
  \bibfield  {author} {\bibinfo {author} {\bibfnamefont {V.~M.}\ \bibnamefont
  {Edelstein}},\ }\href {\doibase 10.1103/PhysRevB.67.020505} {\bibfield
  {journal} {\bibinfo  {journal} {Phys. Rev. B}\ }\textbf {\bibinfo {volume}
  {{\bf 67}}},\ \bibinfo {pages} {020505} (\bibinfo {year} {2003})}\BibitemShut
  {NoStop}%
\bibitem [{\citenamefont {Rashba}(2012)}]{Rashba:2012}%
  \BibitemOpen
  \bibfield  {author} {\bibinfo {author} {\bibfnamefont {E.~I.}\ \bibnamefont
  {Rashba}},\ }\href {\doibase 10.1103/PhysRevB.86.125319} {\bibfield
  {journal} {\bibinfo  {journal} {Phys. Rev. B}\ }\textbf {\bibinfo {volume}
  {86}},\ \bibinfo {pages} {125319} (\bibinfo {year} {2012})}\BibitemShut
  {NoStop}%
\bibitem [{\citenamefont {LaShell}\ \emph {et~al.}(1996)\citenamefont
  {LaShell}, \citenamefont {McDougall},\ and\ \citenamefont
  {Jensen}}]{LaShell:1996}%
  \BibitemOpen
  \bibfield  {author} {\bibinfo {author} {\bibfnamefont {S.}~\bibnamefont
  {LaShell}}, \bibinfo {author} {\bibfnamefont {B.~A.}\ \bibnamefont
  {McDougall}}, \ and\ \bibinfo {author} {\bibfnamefont {E.}~\bibnamefont
  {Jensen}},\ }\href {\doibase 10.1103/PhysRevLett.77.3419} {\bibfield
  {journal} {\bibinfo  {journal} {Phys. Rev. Lett.}\ }\textbf {\bibinfo
  {volume} {{\bf 77}}},\ \bibinfo {pages} {3419} (\bibinfo {year}
  {1996})}\BibitemShut {NoStop}%
\bibitem [{\citenamefont {Koroteev}\ \emph {et~al.}(2004)\citenamefont
  {Koroteev}, \citenamefont {Bihlmayer}, \citenamefont {Gayone}, \citenamefont
  {Chulkov}, \citenamefont {Bl\"ugel}, \citenamefont {Echenique},\ and\
  \citenamefont {Hofmann}}]{Koroteev:2004}%
  \BibitemOpen
  \bibfield  {author} {\bibinfo {author} {\bibfnamefont {Y.~M.}\ \bibnamefont
  {Koroteev}}, \bibinfo {author} {\bibfnamefont {G.}~\bibnamefont {Bihlmayer}},
  \bibinfo {author} {\bibfnamefont {J.~E.}\ \bibnamefont {Gayone}}, \bibinfo
  {author} {\bibfnamefont {E.~V.}\ \bibnamefont {Chulkov}}, \bibinfo {author}
  {\bibfnamefont {S.}~\bibnamefont {Bl\"ugel}}, \bibinfo {author}
  {\bibfnamefont {P.~M.}\ \bibnamefont {Echenique}}, \ and\ \bibinfo {author}
  {\bibfnamefont {P.}~\bibnamefont {Hofmann}},\ }\href {\doibase
  10.1103/PhysRevLett.93.046403} {\bibfield  {journal} {\bibinfo  {journal}
  {Phys. Rev. Lett.}\ }\textbf {\bibinfo {volume} {{\bf 93}}},\ \bibinfo
  {pages} {046403} (\bibinfo {year} {2004})}\BibitemShut {NoStop}%
\bibitem [{\citenamefont {van Weperen}\ \emph {et~al.}(2015)\citenamefont {van
  Weperen}, \citenamefont {Tarasinski}, \citenamefont {Eeltink}, \citenamefont
  {Pribiag}, \citenamefont {Plissard}, \citenamefont {Bakkers}, \citenamefont
  {Kouwenhoven},\ and\ \citenamefont {Wimmer}}]{vanWeperen:2015}%
  \BibitemOpen
  \bibfield  {author} {\bibinfo {author} {\bibfnamefont {I.}~\bibnamefont {van
  Weperen}}, \bibinfo {author} {\bibfnamefont {B.}~\bibnamefont {Tarasinski}},
  \bibinfo {author} {\bibfnamefont {D.}~\bibnamefont {Eeltink}}, \bibinfo
  {author} {\bibfnamefont {V.~S.}\ \bibnamefont {Pribiag}}, \bibinfo {author}
  {\bibfnamefont {S.~R.}\ \bibnamefont {Plissard}}, \bibinfo {author}
  {\bibfnamefont {E.~P. A.~M.}\ \bibnamefont {Bakkers}}, \bibinfo {author}
  {\bibfnamefont {L.~P.}\ \bibnamefont {Kouwenhoven}}, \ and\ \bibinfo {author}
  {\bibfnamefont {M.}~\bibnamefont {Wimmer}},\ }\href {\doibase
  10.1103/PhysRevB.91.201413} {\bibfield  {journal} {\bibinfo  {journal} {Phys.
  Rev. B}\ }\textbf {\bibinfo {volume} {{\bf 91}}},\ \bibinfo {pages} {201413}
  (\bibinfo {year} {2015})}\BibitemShut {NoStop}%
\bibitem [{\citenamefont {Ishizaka}\ \emph {et~al.}(2011)\citenamefont
  {Ishizaka}, \citenamefont {Bahramy}, \citenamefont {Murakawa}, \citenamefont
  {Sakano}, \citenamefont {Shimojima}, \citenamefont {Sonobe}, \citenamefont
  {Koizumi}, \citenamefont {Shin}, \citenamefont {Miyahara}, \citenamefont
  {Kimura}, \citenamefont {Miyamoto}, \citenamefont {Okuda}, \citenamefont
  {Namatame}, \citenamefont {Taniguchi}, \citenamefont {Arita}, \citenamefont
  {Nagaosa}, \citenamefont {Kobayashi}, \citenamefont {Murakami}, \citenamefont
  {Kumai}, \citenamefont {Kaneko}, \citenamefont {Onose},\ and\ \citenamefont
  {Tokura}}]{Ishizaka:2011}%
  \BibitemOpen
  \bibfield  {author} {\bibinfo {author} {\bibfnamefont {K.}~\bibnamefont
  {Ishizaka}}, \bibinfo {author} {\bibfnamefont {M.~S.}\ \bibnamefont
  {Bahramy}}, \bibinfo {author} {\bibfnamefont {H.}~\bibnamefont {Murakawa}},
  \bibinfo {author} {\bibfnamefont {M.}~\bibnamefont {Sakano}}, \bibinfo
  {author} {\bibfnamefont {T.}~\bibnamefont {Shimojima}}, \bibinfo {author}
  {\bibfnamefont {T.}~\bibnamefont {Sonobe}}, \bibinfo {author} {\bibfnamefont
  {K.}~\bibnamefont {Koizumi}}, \bibinfo {author} {\bibfnamefont
  {S.}~\bibnamefont {Shin}}, \bibinfo {author} {\bibfnamefont {H.}~\bibnamefont
  {Miyahara}}, \bibinfo {author} {\bibfnamefont {A.}~\bibnamefont {Kimura}},
  \bibinfo {author} {\bibfnamefont {K.}~\bibnamefont {Miyamoto}}, \bibinfo
  {author} {\bibfnamefont {T.}~\bibnamefont {Okuda}}, \bibinfo {author}
  {\bibfnamefont {H.}~\bibnamefont {Namatame}}, \bibinfo {author}
  {\bibfnamefont {M.}~\bibnamefont {Taniguchi}}, \bibinfo {author}
  {\bibfnamefont {R.}~\bibnamefont {Arita}}, \bibinfo {author} {\bibfnamefont
  {N.}~\bibnamefont {Nagaosa}}, \bibinfo {author} {\bibfnamefont
  {K.}~\bibnamefont {Kobayashi}}, \bibinfo {author} {\bibfnamefont
  {Y.}~\bibnamefont {Murakami}}, \bibinfo {author} {\bibfnamefont
  {R.}~\bibnamefont {Kumai}}, \bibinfo {author} {\bibfnamefont
  {Y.}~\bibnamefont {Kaneko}}, \bibinfo {author} {\bibfnamefont
  {Y.}~\bibnamefont {Onose}}, \ and\ \bibinfo {author} {\bibfnamefont
  {Y.}~\bibnamefont {Tokura}},\ }\href {http://dx.doi.org/10.1038/nmat3051}
  {\bibfield  {journal} {\bibinfo  {journal} {Nature Materials}\ }\textbf
  {\bibinfo {volume} {{\bf 10}}},\ \bibinfo {pages} {521} (\bibinfo {year}
  {2011})}\BibitemShut {NoStop}%
\bibitem [{\citenamefont {Eremeev}\ \emph {et~al.}(2012)\citenamefont
  {Eremeev}, \citenamefont {Nechaev}, \citenamefont {Koroteev}, \citenamefont
  {Echenique},\ and\ \citenamefont {Chulkov}}]{Eremeev:2012}%
  \BibitemOpen
  \bibfield  {author} {\bibinfo {author} {\bibfnamefont {S.~V.}\ \bibnamefont
  {Eremeev}}, \bibinfo {author} {\bibfnamefont {I.~A.}\ \bibnamefont
  {Nechaev}}, \bibinfo {author} {\bibfnamefont {Y.~M.}\ \bibnamefont
  {Koroteev}}, \bibinfo {author} {\bibfnamefont {P.~M.}\ \bibnamefont
  {Echenique}}, \ and\ \bibinfo {author} {\bibfnamefont {E.~V.}\ \bibnamefont
  {Chulkov}},\ }\href {\doibase 10.1103/PhysRevLett.108.246802} {\bibfield
  {journal} {\bibinfo  {journal} {Phys. Rev. Lett.}\ }\textbf {\bibinfo
  {volume} {{\bf 108}}},\ \bibinfo {pages} {246802} (\bibinfo {year}
  {2012})}\BibitemShut {NoStop}%
\bibitem [{\citenamefont {Mourik}\ \emph {et~al.}(2012)\citenamefont {Mourik},
  \citenamefont {Zuo}, \citenamefont {Frolov}, \citenamefont {Plissard},
  \citenamefont {Bakkers},\ and\ \citenamefont {Kouwenhoven}}]{Mourik:2012}%
  \BibitemOpen
  \bibfield  {author} {\bibinfo {author} {\bibfnamefont {V.}~\bibnamefont
  {Mourik}}, \bibinfo {author} {\bibfnamefont {K.}~\bibnamefont {Zuo}},
  \bibinfo {author} {\bibfnamefont {S.~M.}\ \bibnamefont {Frolov}}, \bibinfo
  {author} {\bibfnamefont {S.~R.}\ \bibnamefont {Plissard}}, \bibinfo {author}
  {\bibfnamefont {E.~P. A.~M.}\ \bibnamefont {Bakkers}}, \ and\ \bibinfo
  {author} {\bibfnamefont {L.~P.}\ \bibnamefont {Kouwenhoven}},\ }\href
  {\doibase 10.1126/science.1222360} {\bibfield  {journal} {\bibinfo  {journal}
  {Science}\ }\textbf {\bibinfo {volume} {{\bf 336}}},\ \bibinfo {pages} {1003}
  (\bibinfo {year} {2012})}\BibitemShut {NoStop}%
\bibitem [{\citenamefont {Andreev}(1964)}]{Andreev:1964}%
  \BibitemOpen
  \bibfield  {author} {\bibinfo {author} {\bibfnamefont {A.~F.}\ \bibnamefont
  {Andreev}},\ }\href@noop {} {\bibfield  {journal} {\bibinfo  {journal} {Sov.
  Phys. JETP}\ }\textbf {\bibinfo {volume} {{\bf 19}}},\ \bibinfo {pages}
  {1228} (\bibinfo {year} {1964})}\BibitemShut {NoStop}%
\bibitem [{\citenamefont {Blonder}\ \emph {et~al.}(1982)\citenamefont
  {Blonder}, \citenamefont {Tinkham},\ and\ \citenamefont
  {Klapwijk}}]{Blonder:1982}%
  \BibitemOpen
  \bibfield  {author} {\bibinfo {author} {\bibfnamefont {G.~E.}\ \bibnamefont
  {Blonder}}, \bibinfo {author} {\bibfnamefont {M.}~\bibnamefont {Tinkham}}, \
  and\ \bibinfo {author} {\bibfnamefont {T.~M.}\ \bibnamefont {Klapwijk}},\
  }\href {\doibase 10.1103/PhysRevB.25.4515} {\bibfield  {journal} {\bibinfo
  {journal} {Phys. Rev. B}\ }\textbf {\bibinfo {volume} {{\bf 25}}},\ \bibinfo
  {pages} {4515} (\bibinfo {year} {1982})}\BibitemShut {NoStop}%
\bibitem [{\citenamefont {de~Jong}\ and\ \citenamefont
  {Beenakker}(1995)}]{deJong:1995}%
  \BibitemOpen
  \bibfield  {author} {\bibinfo {author} {\bibfnamefont {M.~J.~M.}\
  \bibnamefont {de~Jong}}\ and\ \bibinfo {author} {\bibfnamefont {C.~W.~J.}\
  \bibnamefont {Beenakker}},\ }\href {\doibase 10.1103/PhysRevLett.74.1657}
  {\bibfield  {journal} {\bibinfo  {journal} {Phys. Rev. Lett.}\ }\textbf
  {\bibinfo {volume} {{\bf 74}}},\ \bibinfo {pages} {1657} (\bibinfo {year}
  {1995})}\BibitemShut {NoStop}%
\bibitem [{\citenamefont {Yokoyama}\ \emph {et~al.}(2006)\citenamefont
  {Yokoyama}, \citenamefont {Tanaka},\ and\ \citenamefont
  {Inoue}}]{Yokoyama:2006}%
  \BibitemOpen
  \bibfield  {author} {\bibinfo {author} {\bibfnamefont {T.}~\bibnamefont
  {Yokoyama}}, \bibinfo {author} {\bibfnamefont {Y.}~\bibnamefont {Tanaka}}, \
  and\ \bibinfo {author} {\bibfnamefont {J.}~\bibnamefont {Inoue}},\ }\href
  {\doibase 10.1103/PhysRevB.74.035318} {\bibfield  {journal} {\bibinfo
  {journal} {Phys. Rev. B}\ }\textbf {\bibinfo {volume} {{\bf 74}}},\ \bibinfo
  {pages} {035318} (\bibinfo {year} {2006})}\BibitemShut {NoStop}%
\bibitem [{\citenamefont {Sun}\ and\ \citenamefont {Shah}(2015)}]{Sun:2015}%
  \BibitemOpen
  \bibfield  {author} {\bibinfo {author} {\bibfnamefont {K.}~\bibnamefont
  {Sun}}\ and\ \bibinfo {author} {\bibfnamefont {N.}~\bibnamefont {Shah}},\
  }\href {\doibase 10.1103/PhysRevB.91.144508} {\bibfield  {journal} {\bibinfo
  {journal} {Phys. Rev. B}\ }\textbf {\bibinfo {volume} {{\bf 91}}},\ \bibinfo
  {pages} {144508} (\bibinfo {year} {2015})}\BibitemShut {NoStop}%
\bibitem [{\citenamefont {de~Gennes}(1966)}]{deGennes}%
  \BibitemOpen
  \bibfield  {author} {\bibinfo {author} {\bibfnamefont {P.~G.}\ \bibnamefont
  {de~Gennes}},\ }\href@noop {} {\emph {\bibinfo {title} {Superconductivity of
  Metals and Alloys}}}\ (\bibinfo  {publisher} {Benjamin, New York},\ \bibinfo
  {year} {1966})\BibitemShut {NoStop}%
\bibitem [{\citenamefont {Demler}\ \emph {et~al.}(1997)\citenamefont {Demler},
  \citenamefont {Arnold},\ and\ \citenamefont {Beasley}}]{Demler:1997}%
  \BibitemOpen
  \bibfield  {author} {\bibinfo {author} {\bibfnamefont {E.~A.}\ \bibnamefont
  {Demler}}, \bibinfo {author} {\bibfnamefont {G.~B.}\ \bibnamefont {Arnold}},
  \ and\ \bibinfo {author} {\bibfnamefont {M.~R.}\ \bibnamefont {Beasley}},\
  }\href {\doibase 10.1103/PhysRevB.55.15174} {\bibfield  {journal} {\bibinfo
  {journal} {Phys. Rev. B}\ }\textbf {\bibinfo {volume} {{\bf 55}}},\ \bibinfo
  {pages} {15174} (\bibinfo {year} {1997})}\BibitemShut {NoStop}%
\bibitem [{\citenamefont {Mineev}\ and\ \citenamefont
  {Samokhin}(1999)}]{MineevBook}%
  \BibitemOpen
  \bibfield  {author} {\bibinfo {author} {\bibfnamefont {V.~P.}\ \bibnamefont
  {Mineev}}\ and\ \bibinfo {author} {\bibfnamefont {K.~V.}\ \bibnamefont
  {Samokhin}},\ }\href@noop {} {\emph {\bibinfo {title} {Introduction to
  Unconventional Superconductivity}}}\ (\bibinfo  {publisher} {Gordon and
  Breach, London},\ \bibinfo {year} {1999})\BibitemShut {NoStop}%
\bibitem [{\citenamefont {VanGennep}\ \emph {et~al.}(2014)\citenamefont
  {VanGennep}, \citenamefont {Maiti}, \citenamefont {Graf}, \citenamefont
  {Tozer}, \citenamefont {Martin}, \citenamefont {Berger}, \citenamefont
  {Maslov},\ and\ \citenamefont {Hamlin}}]{VanGennep:2014}%
  \BibitemOpen
  \bibfield  {author} {\bibinfo {author} {\bibfnamefont {D.}~\bibnamefont
  {VanGennep}}, \bibinfo {author} {\bibfnamefont {S.}~\bibnamefont {Maiti}},
  \bibinfo {author} {\bibfnamefont {D.}~\bibnamefont {Graf}}, \bibinfo {author}
  {\bibfnamefont {S.~W.}\ \bibnamefont {Tozer}}, \bibinfo {author}
  {\bibfnamefont {C.}~\bibnamefont {Martin}}, \bibinfo {author} {\bibfnamefont
  {H.}~\bibnamefont {Berger}}, \bibinfo {author} {\bibfnamefont {D.~L.}\
  \bibnamefont {Maslov}}, \ and\ \bibinfo {author} {\bibfnamefont {J.~J.}\
  \bibnamefont {Hamlin}},\ }\href
  {http://stacks.iop.org/0953-8984/26/i=34/a=342202} {\bibfield  {journal}
  {\bibinfo  {journal} {J. Phys.: Condens. Matter}\ }\textbf {\bibinfo {volume}
  {{\bf 26}}},\ \bibinfo {pages} {342202} (\bibinfo {year} {2014})}\BibitemShut
  {NoStop}%
\bibitem [{\citenamefont {Moroz}\ and\ \citenamefont
  {Barnes}(1999)}]{Moroz:1999}%
  \BibitemOpen
  \bibfield  {author} {\bibinfo {author} {\bibfnamefont {A.~V.}\ \bibnamefont
  {Moroz}}\ and\ \bibinfo {author} {\bibfnamefont {C.~H.~W.}\ \bibnamefont
  {Barnes}},\ }\href {\doibase 10.1103/PhysRevB.60.14272} {\bibfield  {journal}
  {\bibinfo  {journal} {Phys. Rev. B}\ }\textbf {\bibinfo {volume} {60}},\
  \bibinfo {pages} {14272} (\bibinfo {year} {1999})}\BibitemShut {NoStop}%
\bibitem [{\citenamefont {Miller}\ \emph {et~al.}(2003)\citenamefont {Miller},
  \citenamefont {Zumb\"uhl}, \citenamefont {Marcus}, \citenamefont
  {Lyanda-Geller}, \citenamefont {Goldhaber-Gordon}, \citenamefont {Campman},\
  and\ \citenamefont {Gossard}}]{Miller:2003}%
  \BibitemOpen
  \bibfield  {author} {\bibinfo {author} {\bibfnamefont {J.~B.}\ \bibnamefont
  {Miller}}, \bibinfo {author} {\bibfnamefont {D.~M.}\ \bibnamefont
  {Zumb\"uhl}}, \bibinfo {author} {\bibfnamefont {C.~M.}\ \bibnamefont
  {Marcus}}, \bibinfo {author} {\bibfnamefont {Y.~B.}\ \bibnamefont
  {Lyanda-Geller}}, \bibinfo {author} {\bibfnamefont {D.}~\bibnamefont
  {Goldhaber-Gordon}}, \bibinfo {author} {\bibfnamefont {K.}~\bibnamefont
  {Campman}}, \ and\ \bibinfo {author} {\bibfnamefont {A.~C.}\ \bibnamefont
  {Gossard}},\ }\href {\doibase 10.1103/PhysRevLett.90.076807} {\bibfield
  {journal} {\bibinfo  {journal} {Phys. Rev. Lett.}\ }\textbf {\bibinfo
  {volume} {{\bf 90}}},\ \bibinfo {pages} {076807} (\bibinfo {year}
  {2003})}\BibitemShut {NoStop}%
\bibitem [{\citenamefont {Grundler}(2000)}]{Grundler:2000}%
  \BibitemOpen
  \bibfield  {author} {\bibinfo {author} {\bibfnamefont {D.}~\bibnamefont
  {Grundler}},\ }\href {\doibase 10.1103/PhysRevLett.84.6074} {\bibfield
  {journal} {\bibinfo  {journal} {Phys. Rev. Lett.}\ }\textbf {\bibinfo
  {volume} {{\bf 84}}},\ \bibinfo {pages} {6074} (\bibinfo {year}
  {2000})}\BibitemShut {NoStop}%
\bibitem [{\citenamefont {Khodaparast}\ \emph {et~al.}(2004)\citenamefont
  {Khodaparast}, \citenamefont {Doezema}, \citenamefont {Chung}, \citenamefont
  {Goldammer},\ and\ \citenamefont {Santos}}]{Khodaparast:2004}%
  \BibitemOpen
  \bibfield  {author} {\bibinfo {author} {\bibfnamefont {G.~A.}\ \bibnamefont
  {Khodaparast}}, \bibinfo {author} {\bibfnamefont {R.~E.}\ \bibnamefont
  {Doezema}}, \bibinfo {author} {\bibfnamefont {S.~J.}\ \bibnamefont {Chung}},
  \bibinfo {author} {\bibfnamefont {K.~J.}\ \bibnamefont {Goldammer}}, \ and\
  \bibinfo {author} {\bibfnamefont {M.~B.}\ \bibnamefont {Santos}},\ }\href
  {\doibase 10.1103/PhysRevB.70.155322} {\bibfield  {journal} {\bibinfo
  {journal} {Phys. Rev. B}\ }\textbf {\bibinfo {volume} {{\bf 70}}},\ \bibinfo
  {pages} {155322} (\bibinfo {year} {2004})}\BibitemShut {NoStop}%
\bibitem [{\citenamefont {Gilbertson}\ \emph {et~al.}(2009)\citenamefont
  {Gilbertson}, \citenamefont {Branford}, \citenamefont {Fearn}, \citenamefont
  {Buckle}, \citenamefont {Buckle}, \citenamefont {Ashley},\ and\ \citenamefont
  {Cohen}}]{Gilbertson:2009}%
  \BibitemOpen
  \bibfield  {author} {\bibinfo {author} {\bibfnamefont {A.~M.}\ \bibnamefont
  {Gilbertson}}, \bibinfo {author} {\bibfnamefont {W.~R.}\ \bibnamefont
  {Branford}}, \bibinfo {author} {\bibfnamefont {M.}~\bibnamefont {Fearn}},
  \bibinfo {author} {\bibfnamefont {L.}~\bibnamefont {Buckle}}, \bibinfo
  {author} {\bibfnamefont {P.~D.}\ \bibnamefont {Buckle}}, \bibinfo {author}
  {\bibfnamefont {T.}~\bibnamefont {Ashley}}, \ and\ \bibinfo {author}
  {\bibfnamefont {L.~F.}\ \bibnamefont {Cohen}},\ }\href {\doibase
  10.1103/PhysRevB.79.235333} {\bibfield  {journal} {\bibinfo  {journal} {Phys.
  Rev. B}\ }\textbf {\bibinfo {volume} {{\bf 79}}},\ \bibinfo {pages} {235333}
  (\bibinfo {year} {2009})}\BibitemShut {NoStop}%
\bibitem [{\citenamefont {Leontiadou}\ \emph {et~al.}(2011)\citenamefont
  {Leontiadou}, \citenamefont {Litvinenko}, \citenamefont {Gilbertson},
  \citenamefont {Pidgeon}, \citenamefont {Branford}, \citenamefont {Cohen},
  \citenamefont {Fearn}, \citenamefont {Ashley}, \citenamefont {Emeny},
  \citenamefont {Murdin},\ and\ \citenamefont {Clowes}}]{Leontiadou:2011}%
  \BibitemOpen
  \bibfield  {author} {\bibinfo {author} {\bibfnamefont {M.~A.}\ \bibnamefont
  {Leontiadou}}, \bibinfo {author} {\bibfnamefont {K.~L.}\ \bibnamefont
  {Litvinenko}}, \bibinfo {author} {\bibfnamefont {A.~M.}\ \bibnamefont
  {Gilbertson}}, \bibinfo {author} {\bibfnamefont {C.~R.}\ \bibnamefont
  {Pidgeon}}, \bibinfo {author} {\bibfnamefont {W.~R.}\ \bibnamefont
  {Branford}}, \bibinfo {author} {\bibfnamefont {L.~F.}\ \bibnamefont {Cohen}},
  \bibinfo {author} {\bibfnamefont {M.}~\bibnamefont {Fearn}}, \bibinfo
  {author} {\bibfnamefont {T.}~\bibnamefont {Ashley}}, \bibinfo {author}
  {\bibfnamefont {M.~T.}\ \bibnamefont {Emeny}}, \bibinfo {author}
  {\bibfnamefont {B.~N.}\ \bibnamefont {Murdin}}, \ and\ \bibinfo {author}
  {\bibfnamefont {S.~K.}\ \bibnamefont {Clowes}},\ }\href
  {http://stacks.iop.org/0953-8984/23/i=3/a=035801} {\bibfield  {journal}
  {\bibinfo  {journal} {J. Phys.: Condens. Matter}\ }\textbf {\bibinfo {volume}
  {{\bf 23}}},\ \bibinfo {pages} {035801} (\bibinfo {year} {2011})}\BibitemShut
  {NoStop}%
\bibitem [{\citenamefont {Hao}\ \emph {et~al.}(2010)\citenamefont {Hao},
  \citenamefont {Tu}, \citenamefont {Cao}, \citenamefont {Zhou}, \citenamefont
  {Li}, \citenamefont {Guo}, \citenamefont {Fung}, \citenamefont {Ji},
  \citenamefont {Guo},\ and\ \citenamefont {Lu}}]{Hao:2010}%
  \BibitemOpen
  \bibfield  {author} {\bibinfo {author} {\bibfnamefont {X.-J.}\ \bibnamefont
  {Hao}}, \bibinfo {author} {\bibfnamefont {T.}~\bibnamefont {Tu}}, \bibinfo
  {author} {\bibfnamefont {G.}~\bibnamefont {Cao}}, \bibinfo {author}
  {\bibfnamefont {C.}~\bibnamefont {Zhou}}, \bibinfo {author} {\bibfnamefont
  {H.-O.}\ \bibnamefont {Li}}, \bibinfo {author} {\bibfnamefont {G.-C.}\
  \bibnamefont {Guo}}, \bibinfo {author} {\bibfnamefont {W.~Y.}\ \bibnamefont
  {Fung}}, \bibinfo {author} {\bibfnamefont {Z.}~\bibnamefont {Ji}}, \bibinfo
  {author} {\bibfnamefont {G.-P.}\ \bibnamefont {Guo}}, \ and\ \bibinfo
  {author} {\bibfnamefont {W.}~\bibnamefont {Lu}},\ }\href {\doibase
  10.1021/nl101181e} {\bibfield  {journal} {\bibinfo  {journal} {Nano Letters}\
  }\textbf {\bibinfo {volume} {{\bf 10}}},\ \bibinfo {pages} {2956} (\bibinfo
  {year} {2010})}\BibitemShut {NoStop}%
\bibitem [{\citenamefont {Liang}\ and\ \citenamefont {Gao}(2012)}]{Liang:2012}%
  \BibitemOpen
  \bibfield  {author} {\bibinfo {author} {\bibfnamefont {D.}~\bibnamefont
  {Liang}}\ and\ \bibinfo {author} {\bibfnamefont {X.~P.}\ \bibnamefont
  {Gao}},\ }\href {\doibase 10.1021/nl301325h} {\bibfield  {journal} {\bibinfo
  {journal} {Nano Letters}\ }\textbf {\bibinfo {volume} {{\bf 12}}},\ \bibinfo
  {pages} {3263} (\bibinfo {year} {2012})}\BibitemShut {NoStop}%
\bibitem [{\citenamefont {Dyakonov}\ and\ \citenamefont
  {Perel}(1972)}]{Dyakonov:1972}%
  \BibitemOpen
  \bibfield  {author} {\bibinfo {author} {\bibfnamefont {M.~I.}\ \bibnamefont
  {Dyakonov}}\ and\ \bibinfo {author} {\bibfnamefont {V.~I.}\ \bibnamefont
  {Perel}},\ }\href@noop {} {\bibfield  {journal} {\bibinfo  {journal} {Sov.
  Phys. Solid State}\ }\textbf {\bibinfo {volume} {{\bf 13}}},\ \bibinfo
  {pages} {3023} (\bibinfo {year} {1972})}\BibitemShut {NoStop}%
\bibitem [{\citenamefont {Yi}\ \emph {et~al.}(2015)\citenamefont {Yi},
  \citenamefont {Kiselev}, \citenamefont {Thorp}, \citenamefont {Noah},
  \citenamefont {Nguyen}, \citenamefont {Bui}, \citenamefont {Rajavel},
  \citenamefont {Hussain}, \citenamefont {Gyure}, \citenamefont {Kratz},
  \citenamefont {Qian}, \citenamefont {Manfra}, \citenamefont {Pribiag},
  \citenamefont {Kouwenhoven}, \citenamefont {Marcus},\ and\ \citenamefont
  {Sokolich}}]{Yi:2015}%
  \BibitemOpen
  \bibfield  {author} {\bibinfo {author} {\bibfnamefont {W.}~\bibnamefont
  {Yi}}, \bibinfo {author} {\bibfnamefont {A.~A.}\ \bibnamefont {Kiselev}},
  \bibinfo {author} {\bibfnamefont {J.}~\bibnamefont {Thorp}}, \bibinfo
  {author} {\bibfnamefont {R.}~\bibnamefont {Noah}}, \bibinfo {author}
  {\bibfnamefont {B.-M.}\ \bibnamefont {Nguyen}}, \bibinfo {author}
  {\bibfnamefont {S.}~\bibnamefont {Bui}}, \bibinfo {author} {\bibfnamefont
  {R.~D.}\ \bibnamefont {Rajavel}}, \bibinfo {author} {\bibfnamefont
  {T.}~\bibnamefont {Hussain}}, \bibinfo {author} {\bibfnamefont {M.~F.}\
  \bibnamefont {Gyure}}, \bibinfo {author} {\bibfnamefont {P.}~\bibnamefont
  {Kratz}}, \bibinfo {author} {\bibfnamefont {Q.}~\bibnamefont {Qian}},
  \bibinfo {author} {\bibfnamefont {M.~J.}\ \bibnamefont {Manfra}}, \bibinfo
  {author} {\bibfnamefont {V.~S.}\ \bibnamefont {Pribiag}}, \bibinfo {author}
  {\bibfnamefont {L.~P.}\ \bibnamefont {Kouwenhoven}}, \bibinfo {author}
  {\bibfnamefont {C.~M.}\ \bibnamefont {Marcus}}, \ and\ \bibinfo {author}
  {\bibfnamefont {M.}~\bibnamefont {Sokolich}},\ }\href {\doibase
  http://dx.doi.org/10.1063/1.4917027} {\bibfield  {journal} {\bibinfo
  {journal} {Applied Physics Letters}\ }\textbf {\bibinfo {volume} {{\bf
  106}}},\ \bibinfo {eid} {142103} (\bibinfo {year} {2015})}\BibitemShut
  {NoStop}%
\end{thebibliography}%

\end{document}